\documentclass{jfm}
\usepackage{graphicx,color}
\usepackage{float}
\usepackage{epstopdf, epsfig}
\usepackage{caption,subcaption}
\usepackage[intlimits]{amsmath}
\usepackage{lscape}
\usepackage{rotating}

\shorttitle{Breakup of surfactant-laden  liquid sheets}
\shortauthor{C. R. Constante-Amores  et al.}

\title{Role of surfactant-induced Marangoni stresses in retracting  liquid sheets
}

\author{C. R. Constante-Amores$^{1,2}$, \ns J. Chergui$^3$, \ns S. Shin$^4$,  \ns D. Juric$^{3,5}$,
\ns J. R. Castrej\'on-Pita$^{6}$,
\ns A. A. Castrej\'on-Pita$^{1}$\corresp{\email{alfonso.castrejon-pita@wadham.ox.ac.uk}}%
}
  
\affiliation{
$^1$Department of Engineering Science, University of Oxford, Oxford OX1 3PJ, UK\\[\affilskip]
$^2$ Mathematical Institute, University of Oxford, Oxford OX2 6GG, UK\\[\affilskip]
$^3$ Universit\'e Paris Saclay, Centre National de la Recherche Scientifique (CNRS), Laboratoire Interdisciplinaire des Sciences du Num\'erique (LISN), 91400 Orsay, France
\\[\affilskip]
$^4$Department of Mechanical and System Design Engineering, Hongik University, S. Korea
\\[\affilskip]
$^5$Department of Applied Mathematics and Theoretical Physics, University of Cambridge,  Cambridge CB3 0WA, UK
\\[\affilskip]
$^6$School of Engineering and Material Science, Queen Mary University of London, Mile End Road, London E1 4NS, UK

}

\begin{document}

\maketitle

\begin{abstract}

In this work, we study the effect of insoluble surfactants on the three-dimensional rim-driven retraction dynamics of thin water sheets in air. We employ an interface-tracking/level-set method to ensure the full coupling between the surfactant-induced Marangoni-stresses, interfacial diffusion, and inertia. 
Our findings are contrasted with the (Newtonian) dynamics of a liquid sheet edge finding that the surfactant concentration can delay, or effectively prevent, the breakup of the rim.
Our simulations use the fastest growing Rayleigh-Plateau instability to drive droplet detachment from the fluid sheet (rim).
The results of this work unravel the significant role of Marangoni stresses in the retracting sheet dynamics at large elasticity numbers.  
We study the sensitivity of the dynamics to the elasticity number and the rigidification of the interface.

\end{abstract}

\section{Introduction}

The capillary  break-up of liquid films into droplets plays a major role in various unsteady fluid fragmentation phenomena, such as atomisation or droplet splashing  \citep{villermaux_2020}. It is therefore unsurprising that sheet retraction dynamics have received significant interest since the initial experimental observations by \citet{Dupre_1867} and \citet{Rayleigh_1879}.
Capillary film retraction follows the puncturing of a static liquid film, where a capillary-induced flow drives the opening of the interface (a hole) at a constant speed, i.e. the  Taylor-Culick velocity  \citep{Taylor_1959,Culick}. 
Early pictures by \citet{Rayleigh_1891} and \citet{Ranz} show that the opening of the sheet results in the formation of a rim with roughly cylindrical caps, as the film moves away from the puncture. 
The desire to understand the fundamental mechanism underlying the  development of the hole-driven expansion has led to numerous studies in this field, see for instance \citet{Debregeas_1995,Keller_1995,Brenner_1999,Fullana_1999, Sunderhauf_2002, savva_bush_2009, 
Roisman,Krechetnikov,Lhuissier_2011,Villermaux_2011,Gordillo_2011,Agbaglah_2013,agbaglah_2021}.

\citet{Debregeas_1995} and \citet{savva_bush_2009} 
concluded that the retracting dynamics of a liquid sheet are governed by a balance between inertial, viscous, and surface-tension forces. Accordingly,  the Ohnesorge number, $Oh$, (e.g., ratio of viscous to capillary forces) becomes the most appropriate parameter to parametrise the 
flow dynamics.
Different regimes have been identified that depend on $Oh$:
(i) in the $Oh< 1$ regime, which is characterised by a nearly inviscid flow, the dynamics are driven by surface tension leading to the formation of capillary waves ahead of the roughly cylindrical rim;
(ii) in the $Oh>10$ regime viscous forces dominate the dynamics causing the suppression of the rim;
and
(iii) an intermediate regime bridges the two previous regimes. 

The groundbreaking experiments of \citet{McEntee_1969}  proposed that a surface instability along the rim drives the formation of ligaments, and eventually, the ejection of droplets. 
Several studies have suggested that different physical mechanisms prompt the detachment of these droplets: a Rayleigh-Plateau (RP) instability (discussed below), a nonlinear amplification mechanism \citep{yarin_weiss_1995}, a Richtmyer-Meshkov instability, and a competition/collaboration between the Richtmyer–Meshkov and Rayleigh-Taylor (RT) instabilities \citep{Krechetnikov}.  
However, \citet{bremond_villermaux_2006} (for the fragmentation of the lamella of  colliding  jets), \citet{Rieber} and \citet{Zhang_2010} (for the fragmentation of the sheet-crown  
from impacts of drops onto a thin layer of liquid)  concluded that the RP instability is  responsible for the onset of  drop detachment. 
These findings contrast with the previous study by \citet{Fullana_1999}, that suggest that the growth of the rim during the retraction prevents the onset of a RP instability.
Finally, \citet{Agbaglah_2013} (for a numerical sheet retraction  in a frame of reference moving at the Taylor-Culick velocity) and \citet{Wang_prl} (for drop impact onto a solid surface) 
showed that both  RT and RP instabilities are responsible for droplet generation from the rim. 
The former is important at early times during the deceleration of the ejected sheet, whereas the latter is  predominant at longer times once the rim has already formed.
According to recent studies by \citet{wang_bourouiba_2018,wang_bourouiba_2021}, the ligament dynamics can result in end-pinching,  break up into satellite droplets or recombination into a single fluid volume.

\citet{McEntee_1969} used surfactants in their experimental work to study the bursting of liquid films without directly endorsing the effects of Marangoni-induced flows. 
In contrast, recent studies have demonstrated the crucial role of surfactants on the dynamics of the capillary singularity \citep{Ambravaneswaran,timmermans_lister_2002,craster_pof_2002,Liao_2006, Kamat_prf_2018}. 
 \citet{Constante-Amores_prf_2020} and   \citet{kamat_2020} 
showed that surfactant-induced Marangoni-stresses inhibit end-pinching for retracting liquid threads via the
suppression of stagnation points, which in turn leads to flow reversal near the vicinity of the neck. 

Three-dimensional numerical simulations of retracting liquid sheets are scarce in the literature due to the need for large aspect ratios  
to induce both the growth of instabilities at the rim, and the development of the nonlinear flow dynamics. 
In this work, the  role of surfactant-induced Marangoni stresses on the  retraction of thin liquid sheets and the detachment of droplets is studied in a  three-dimensional, nonlinear framework.
This paper  is structured as follows: Section \ref{Numerical} introduces the numerical method, governing dimensionless parameters, problem configuration, and validation; Section \ref{sec:Results} provides the results from the simulations, and concluding remarks are given in Section \ref{sec:Con}.

\section{Problem formulation and numerical method\label{Numerical}}
%

Numerical simulations were performed by solving the transient two-phase Navier-Stokes equations in a three-dimensional Cartesian domain $\mathbf{x} = \left(x, y, z \right)$ (see figure \ref{configuration}). 
The interface is captured via a  hybrid front-tracking/level-set method; and a 
convective-diffusion equation is solved  for the surfactant transport along the deforming interface \citep{Shin_jcp_2018}. 
All variables are made dimensionless following:
\begin{equation}  \label{scales}
\centering
\quad \tilde{\mathbf{x}}=\frac{\mathbf{x}}{h_0},
\quad \tilde{t}=\frac{t}{t_{inv}}, 
\quad \tilde{\textbf{u}}=\frac{\textbf{u}} {u_{TC}},
\quad \tilde{p}=\frac{p}{\rho_l u_{TC}^2}, 
\quad \tilde{\sigma}=\frac{\sigma}{\sigma_s},
\quad \tilde{\Gamma}=\frac{\Gamma}{\Gamma_\infty},
\end{equation}

\noindent	
where, $t$, $\textbf{u}$, and $p$ stand for time, velocity, and pressure, respectively. The physical parameters correspond to the liquid density $\rho_l$, viscosity, $\mu_l$, surface tension, $\sigma$, surfactant-free surface tension, $\sigma_s$, and
$u_{TC}=\sqrt{2 \sigma_s/ (\rho h_0)}$, is the well-known Taylor-Culick speed; \textcolor{black}{here $h_0$ stands for the sheet thickness.}
In this study we focus on the low-$Oh$ regime; hence, the characteristic time scale is given by the capillary time, $t_{inv}= h_0/u_{TC}=\sqrt{\rho_l h_0^3/ (2 \sigma_s)}$. 
As seen, the interfacial surfactant concentration, $\Gamma$, is also made dimensionless through the saturation interfacial concentration, $\Gamma_{\infty}$. 
\textcolor{black}{As a result of the scaling in equation (\ref{scales}), the dimensionless forms of the governing equations for the flow and the surfactant transport are respectively expressed as}

\begin{equation}\label{div}
\textcolor{black}{ \nabla \cdot \tilde{\textbf{u}}=0,}
\end{equation}
\begin{equation}\label{NS_Eq}
\textcolor{black}{\tilde{\rho} (\frac{\partial \tilde{\textbf{u}}}{\partial \tilde{t}}+\tilde{\textbf{u}} \cdot\nabla \tilde{\textbf{u}}) + \nabla \tilde{p}  =  Oh ~ \nabla\cdot  \left [ \tilde{\mu} (\nabla \tilde{\textbf{u}} +\nabla \tilde{\textbf{u}}^T) \right ] +
\int \limits_{\tilde{A}\tilde{(t)}} \left(\tilde{\sigma} \tilde{\kappa} \textbf{n}  +  \nabla_s  \tilde{\sigma} \right) \delta \left(\tilde{\textbf{x}} - \tilde{\textbf{x}}_{_f}  \right)\mbox{d}\tilde{A},}
\end{equation}

 \begin{equation} 
 \label{equation_surfactant}
\textcolor{black}{ \frac{\partial \tilde{\Gamma}}{\partial \tilde{t}}+\nabla_s \cdot (\tilde{\Gamma} \tilde{\textbf{u}} _{\rm t}) =\frac{1}{Pe_s} \nabla^2_s \tilde{\Gamma},}
 \end{equation}

\noindent
\textcolor{black}{
which correspond to the equations of mass and momentum conservation and the convective-diffusion equations for the interfacial concentration, respectively. 
Here, the density $\tilde{\rho}$ and viscosity $\tilde{\mu}$ are defined by $\tilde{\rho}=\rho_g/\rho_l + \left(1 -\rho_g/\rho_l\right) \mathcal{H}\left(\tilde{\textbf{x}},\tilde{t}\right)$ and $\tilde{\mu}=\mu_g/\mu_l+ \left(1 -\mu_g/\mu_l\right) \mathcal{H}\left( \tilde{\textbf{x}},\tilde{t}\right)$ wherein $\mathcal{H}\left( \tilde{\textbf{x}},\tilde{t}\right)$ represents a smoothed Heaviside function. In this work, $\mathcal{H}\left( \tilde{\textbf{x}},\tilde{t}\right)$ is zero in the gas phase and unity in the liquid. The subscript $g$ designates the gas phase, $\tilde{\textbf{u}}_{\rm{t}}= \left ( \tilde{\textbf{u}}_{\rm{s}} \cdot \textbf{t} \right ) \textbf{t}$ stands for the velocity vector tangential to the interface in which $\tilde{\textbf{u}}_{\rm{s}}$ corresponds to the interfacial velocity,
$\kappa$ stands for the interfacial curvature  obtained from the Lagrangian interface structure, and 
$\nabla_s=\left({\mathbf{I}}-\mathbf{n}\mathbf{n}\right)\cdot \nabla$ stands for the surface gradient operator wherein $\mathbf{I}$ is the identity tensor and $\mathbf{n}$ is the outward-pointing unit normal to the interface. Finally, $\tilde{\mathbf{x}}_f$ is the parameterisation of the interface $\tilde{A} (\tilde{t})$, and 
$\delta$ represents a Dirac delta function that is non-zero when $\tilde{\mathbf{x}}=\tilde{\mathbf{x}}_f$ only.}

The dimensionless groups that govern the dynamics are defined as
\begin{equation}
Oh=\frac{\mu_l}{\sqrt{\sigma_s h_0 \rho_l}}, ~~~
Pe_s=\frac{ u_{TC} h_0}{D_s},     ~~~
\beta_s= \frac{\Re T \Gamma_\infty}{\sigma_s} 
\end{equation}
where $Oh$ and  $Pe_s$ stand for the Ohnesorge and (interfacial) Peclet numbers, respectively, while $\beta_s$ is the surfactant elasticity number which provides a measure of the sensitivity of $\sigma$ to $\Gamma$;
here, 
$\Re$ is the ideal gas constant value ($\Re = 8.314$ J K$^{-1}$ mol$^{-1}$),  $T$ denotes temperature \textcolor{black}{ and $D_s$ stands for the diffusion coefficient}. The non-linear Langmuir equation expresses $\sigma$ in terms of $\Gamma$, i.e. $\tilde{\sigma}=1 + \beta_s \ln{ (1 -\tilde{\Gamma})}$, and the Marangoni stress, $\tilde{\tau}$, is given as a function of $\tilde{\Gamma}$ as $\tilde{\tau}= \nabla_s \tilde{\sigma}\cdot {\mathbf{t}} =-\beta_s/ (1-\tilde{\Gamma})\nabla_s\tilde{\Gamma}\cdot {\mathbf{t}}$. 
Tildes are dropped henceforth. 

\textcolor{black}{We present a brief summary of the numerical method used in this study. The Navier Stokes equations are solved by a finite volume method on a staggered grid (\citet{Harlow_pof_1965}). The computational domain is discretised by a fixed regular grid (i.e. Eulerian grid) and the spatial derivatives are approximated by standard centred difference discretisation, except for the non-linear term, which makes use of a second-order essentially non-oscillatory (ENO) scheme, \citet{Sussman_cp_1998}. The interface is tracked explicitly by an additional Lagrangian grid by using the Front-Tracking method, and the interface is reconstructed  by using the Level Contour Reconstruction Method  \citet{Shin_jcp_2002,Shin_ijnmf_2009}. Communication between the Eulerian and Lagrangian grid (for the transfer of the geometric information of the interface) is done by using the discrete delta function and the Immersed Boundary Method of \citet{Peskin_jcp_1977}. The advection of the Lagrangian interface is done by  integrating $d\textbf{x}_f /dt = \textbf{V}$ with a second-order Runge-Kutta method, where $\textbf{V}$ stands for the interfacial velocity which has been calculated by interpolation from the Eulerian velocity. 
The code is parallelised through an algebraic domain-decomposition technique and the communication between subdomains for data exchange is managed by the Message Passing Interface (MPI) protocol. More details of the numerical method used to solve the above equations
is described in detail by \citet{Shin_jcp_2018}}

\subsection{Numerical setup and validation}

\begin{figure}
\begin{center} 
\begin{tabular}{cc}
 \includegraphics[width=0.49\linewidth]{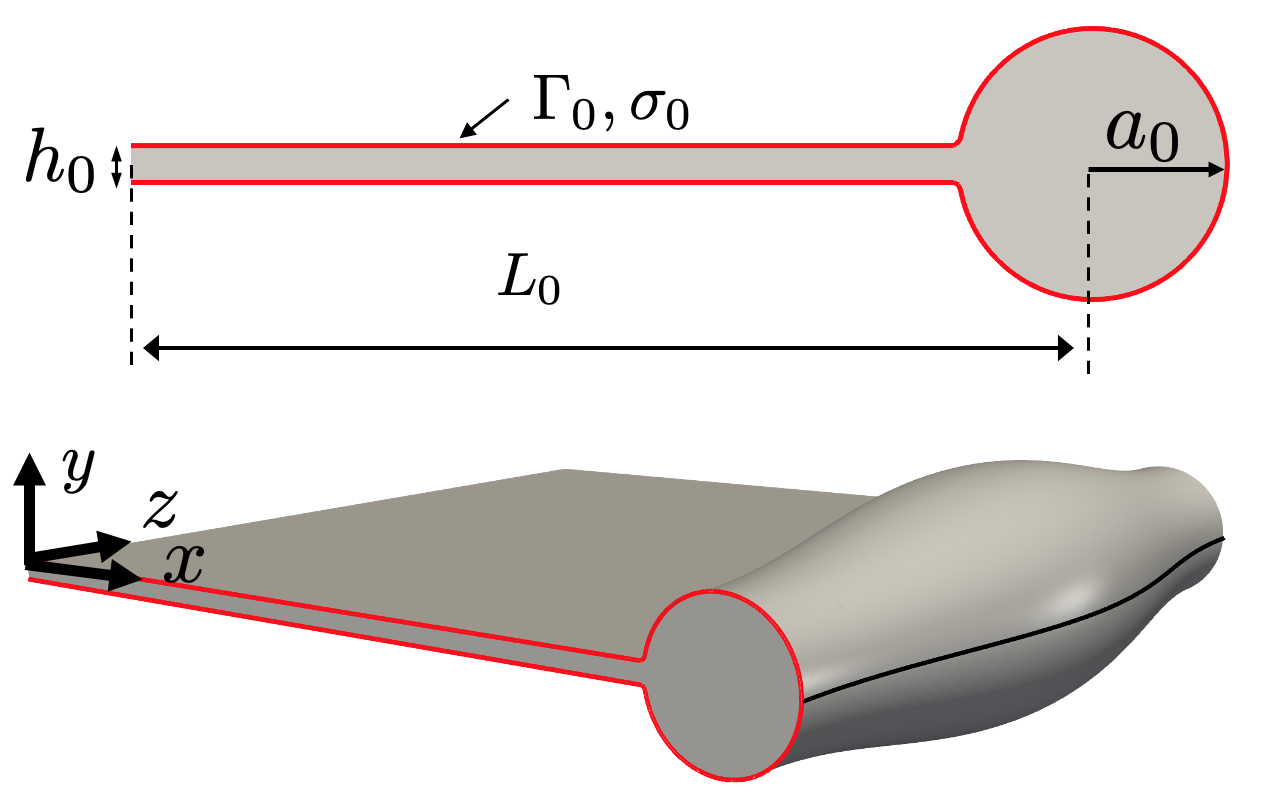}   & 
  \includegraphics[width=0.45\linewidth]{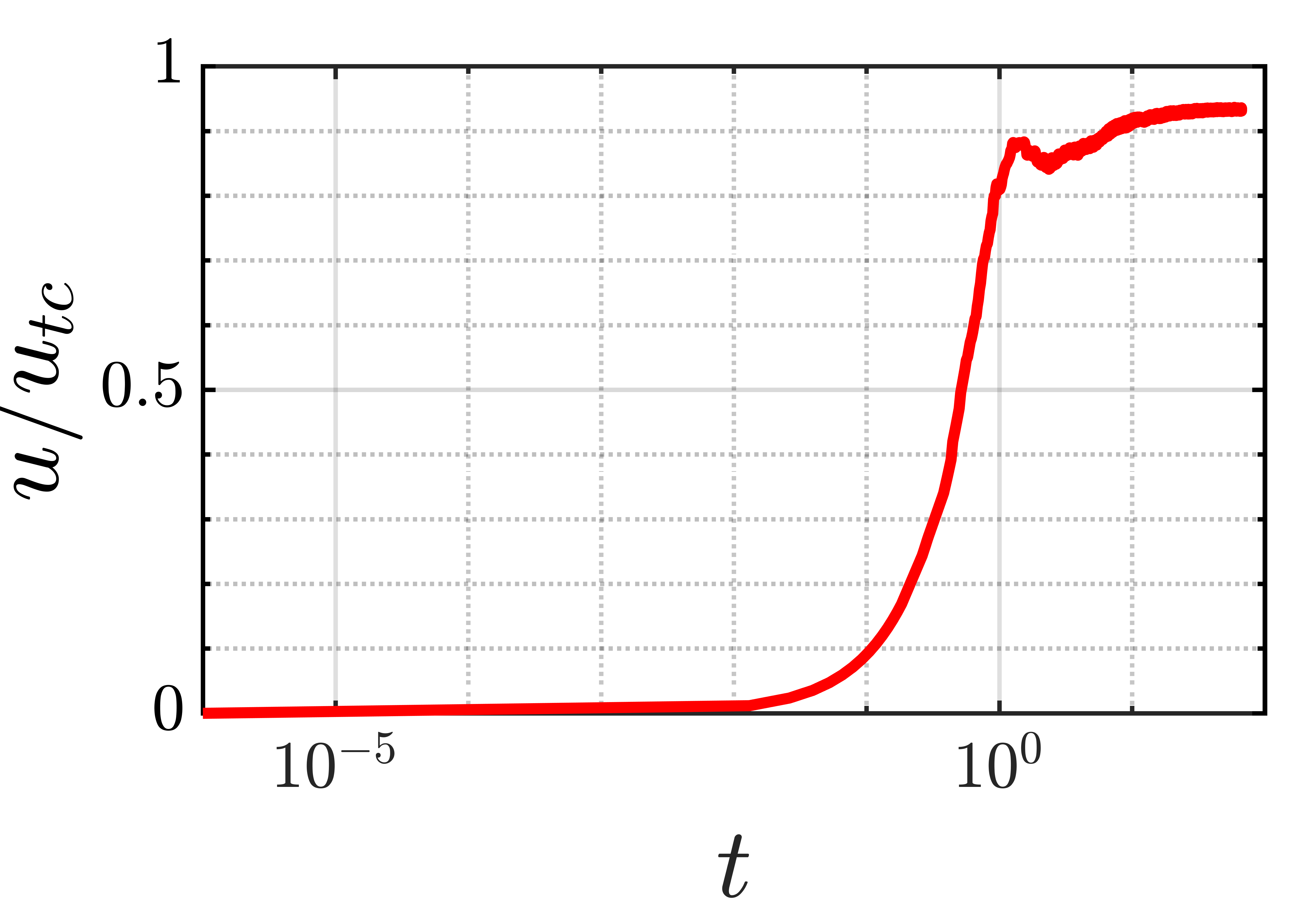}  \\
    (a) & (b) \\
\end{tabular}
\end{center} 
\caption{\label{configuration} (a) Schematic view of the sheet and rim geometry: the initial shape of the retracting sheet (not-to-scale) is characterised by a cylindrical rim with  $e=0.2$ and an initial instability amplitude of $\epsilon=0.25$. The black line represents the arc-length, $s$, across the $x-z$ plane $(y=\lambda/2)$. (b) presents the numerical results for the retracting velocity in terms of the Taylor-Culick value $u_{TC}$ for a surfactant-free liquid sheet with $Oh=0.1$, $\epsilon=0$ and $e=2$.
} \vspace{-0.3 cm}
\end{figure}

Figure \ref{configuration}a illustrates the geometry used in this study, we have initialised the sheet film dynamics using a linearly unstable configuration as proposed by \citet{Agbaglah_2013}. 
Thus, we have considered a thin liquid sheet of thickness $h_0$, and initial length $L_0$, connected to a cylindrical rim of radius $a_0$.
The initial shape of the rim  is given by 
$a(\omega,\varepsilon)=a_0 \left [1+\varepsilon \cos\left ( \omega z \right )\right ]$,
where $\varepsilon$ and $\omega$ stand for the perturbation amplitude  and the growth rate, respectively. 
The dispersion relation for the  Rayleigh-Plateau instability suggests  that the largest growth rate, $\omega$, is obtained for a \textcolor{black}{wavenumber} $ \textcolor{black}{k} \sim 0.6970$, which corresponds to $\lambda_{max}=2\pi/\textcolor{black}{k} =9.016 a_0$ (where $\lambda_{max}$ is the most unstable wavelength).
All  simulations were run using an initial perturbation amplitude of $\varepsilon=0.25$ with $\textcolor{black}{k} \sim 0.6970$, unless stated otherwise in the text.

Additionally to the dimensionless flow parameters, the flow dynamics are also controlled by  the ratio of the film thickness to the radius of the rim, i.e.  $e=h_0/a_0$. 
In this study, we consider the zero acceleration limit, and $e=0.2$, following the work by \citet{Agbaglah_2013} \textcolor{black}{ where their  stability analysis and dispersion relationship depended on the initial acceleration.}
As the liquid sheet retracts, the thickness of the rim grows over time as the rim engulfs the sheet,  $e$ decreases, and the rim deceleration vanishes resulting in the rim dynamics being fully driven by the RP instability. 
Therefore, the flow dynamics depend on two competing time scales: the time scale given by the RP instability in the (spanwise) $z$-direction ($T_{RP}$), and the time scales of the rim growth ($T_{ret}$).
The time scale of the action of surface tension and droplet detachment is given by the capillary time, $T_{RP}=\sqrt{\rho_l a_0^3/ \sigma_s}$.
On the other hand, mass conservation results in the scaling $a (da/dt) \sim h u_{TC}$, leading to  $T_{ret}=a/(da/dt) \sim \sqrt{\rho_l a_0^4/(2 \sigma_s h_0)}$; for $T_{RP} <<T_{ret}$, then,  the liquid sheet must satisfy  $\sqrt{h_0/a_0}<<1$,  as $a_0^2 \sim h_0L_0$, which leads to the next condition  $\sqrt[4]{h_0/L_0}<<1$ that is accomplished at  very high aspect ratios (see \citet{Mirjalili_snh_2018} for more details). 
Finally, we justify the initial sinusoidal perturbation by looking into the breakup time $T_B$ as a function of $T_B=f(Oh,\varepsilon)$ (see expression $5$ from \citet{Driessen}); for  $T_B=f(Oh,\varepsilon=0.25) \sim T_{RP}. $

The computational domain corresponds to $6.75  \lambda_{max} \times ~ \lambda_{max} \times ~ \lambda_{max} $ (i.e., $304.29 h_0 \times 45.08h_0  \times 45.08h_0  $) where the $x$-coordinate is aligned to the length  of the sheet. The domain is sufficiently large in the streamwise direction to avoid the effect of artificial reflections from the boundary of this direction. The simulations are initialised with fluids at rest in the absence of gravity. 
The periodic boundary condition is imposed at all the variables in the (spanwise) $z$-direction of the domain. 
A no-penetration boundary condition is prescribed for the bottom and top domain. 
The domain is discretised with a regular cubic mesh of  size $\Delta \textbf{x}= h_0/6$, ensuring that our numerical predictions are mesh independent; our results do not vary with decreasing grid size.
Note that similar mesh sizes have been used in previously studies, \citet{agbaglah_2021} and \citet{Gordillo_2011}, for  three- and two-dimensional simulations of surfactant-free retracting liquid sheets. Additionally, liquid volume and surfactant mass conservation are satisfied with errors under $0.001\%$. Extensive mesh studies for surface-tension-driven phenomena with the same numerical methodology have been presented in \citet{Constante-Amores_prf_2020,constanteamores2020bb,constante_coales}
At the early stages of the simulation, capillary action drives the formation of a smooth cylindrical rim at the free end of the liquid sheet. After a short transient stage, the dynamics reach the Taylor–Culick regime, characterised by a constant retracting velocity.
Figure \ref{configuration}b contrasts our numerical results against the expected Taylor-Culick retraction velocity for $Oh=0.1$, $\epsilon=0$ and $e=2$.
As seen, the  constant retracting velocity is $\sim8\%$ smaller than the  predicted Taylor-Culick velocity. 
This difference is consistent with observation by \citet{agbaglah_2021} and \citet{Song_1999} and are explained by the large density and viscosity ratios.
The retracting dynamics of a liquid thread and non-linear dynamics before pinch-off have been validated previously by \citet{Constante-Amores_prf_2020}. We refer to \citet{Shin_jcp_2018} for the accuracy of the surfactant equations.

\subsection{Parameters}

Our conditions assume a retracting water liquid sheet in air, i.e. $\rho_g /  \rho_l = 1.2 \times 10^{-3}$ and $\mu_g / \mu_l = 0.018 $, as in the previous work of \citet{Agbaglah_2013} and \citet{agbaglah_2021}.
We assume a thickness of a soap film of $h_0 = 2.0 \ \mu$m, i.e. $Oh=0.0833$ (in agreement with \citet{Meister_aiche_1969}).  
The parameter $\beta_s$ depends on $\Gamma_\infty$ and therefore on the critical micelle concentration (CMC), i.e.  $\Gamma_\infty \sim \mathcal{O}(10^{-6})$ mol m$^{-2}$ for NBD-PC (1-palmitoyl-2-12-[(7-nitro-2-1,3-benzoxadiazol-4- yl)amino]dodecanoyl-sn-glycero-3 -phosphocholine); thus, here we explore the range of $0.1 < \beta_s < 0.5$. We have set $Pe_s = 10^2$ following \citet{batchvarov2020effect} 
who suggested that the interfacial dynamics are weakly-dependent on $Pe_s$ beyond this value \textcolor{black}{i.e., convective effects, driven by surface tension gradients (Marangoni stresses), rather than surface diffusion, dominate the interfacial distribution of surfactant when $Pe_s >10^2$}.
Hence, for a liquid sheet 
characterised by $h_0 = 2.0 \ \mu$m, the relevant time scales are given by $T_{RP} \sim \mathcal{O}(10^{-8})$ s, $T_{ret} \sim \mathcal{O}(10^{-6})$ s, and the  Marangoni time-scale, $T_{\tau} \sim \mu_l h_0/ \Delta \sigma \sim \mathcal{O}(10^{-8})$ s. Consequently, Marangoni stresses play a key role in the flow dynamics. 

\section{Results\label{sec:Results}}

\begin{figure}
\begin{center} 
\includegraphics[width=\linewidth]{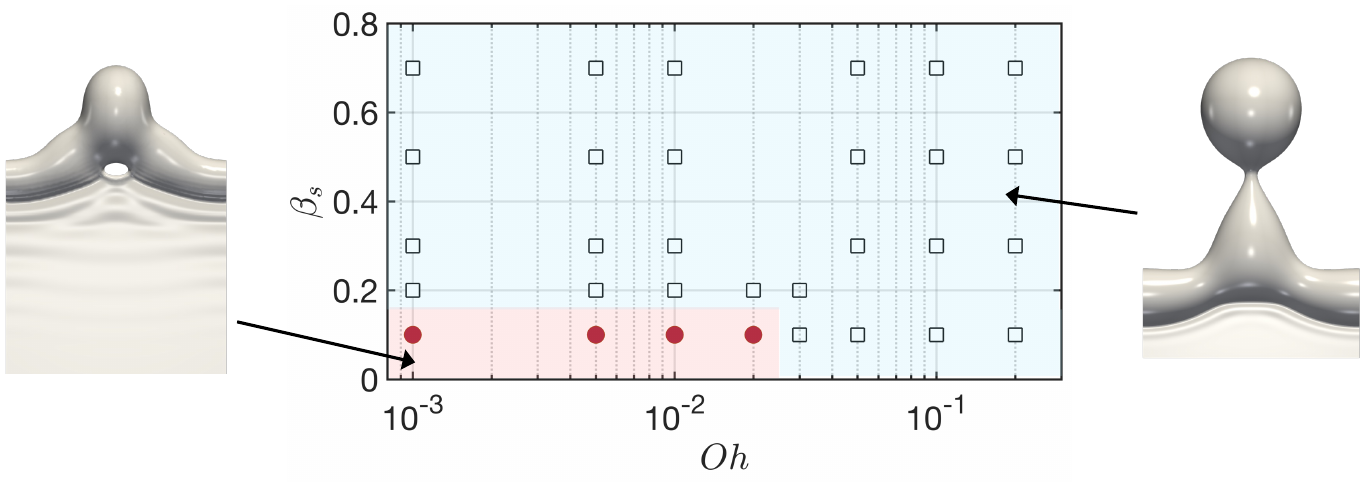}  
\end{center} 
\caption{\label{Regime_map}
\textcolor{black}{Regime map of the phenomenological interfacial dynamics in the $\beta_s$-$Oh$ space. 
Circles represent numerical simulations in which capillary waves induce the formation of a hole behind the rim; hollow square symbols represent a droplet shedding from the rim via a RP-instability. Here, $Pe_s=100$, $\Gamma=\Gamma_\infty/2$, $e=0.2$ and  $\epsilon=0.25$.}
} 
\end{figure}

\textcolor{black}{ We start the discussion of the results by presenting a phenomenological picture of the interfacial dynamics in a phase diagram in a $\beta_s$-$Oh$ space. 
These dynamics range from a nearly inviscid sheet ($Oh \sim 10^{-3}$) to typical soap films ($Oh \sim 10^{-1}$).
Figure \ref{Regime_map} shows the regime map in terms of the interfacial dynamics predicted from our numerical simulations. 
Our results show that two distinct regimes can be identified based on the outcome of the retracting dynamics. 
At low $\beta_s$ and $Oh$, surfactant-driven Marangoni stresses do not promote the reopening of the adjacent sheet connected to the rim; thus, capillary waves result in the formation of a hole behind the rim. This hole grows in the spanwise direction and separates the rim from the sheet. In fact, these dynamics have previously been observed by \citet{Mirjalili_snh_2018} and \citet{constante_jets}.
The rest of the $\beta_s$-$Oh$ space is dominated by surfactant-induced Marangoni stresses which reopen the adjacent sheet to promote droplet detachment via RP instability. Below, we focus on this regime, and provide an extensive explanation of the interfacial dynamics.
}

\begin{figure}
\begin{center} 
\begin{tabular}{ccccc}
\hline
$t=49.49$& $t=63.63$& $t=106.06$& $t=141.42$ & $t=189.50$ \\
\hline
\includegraphics[width=0.15\linewidth]{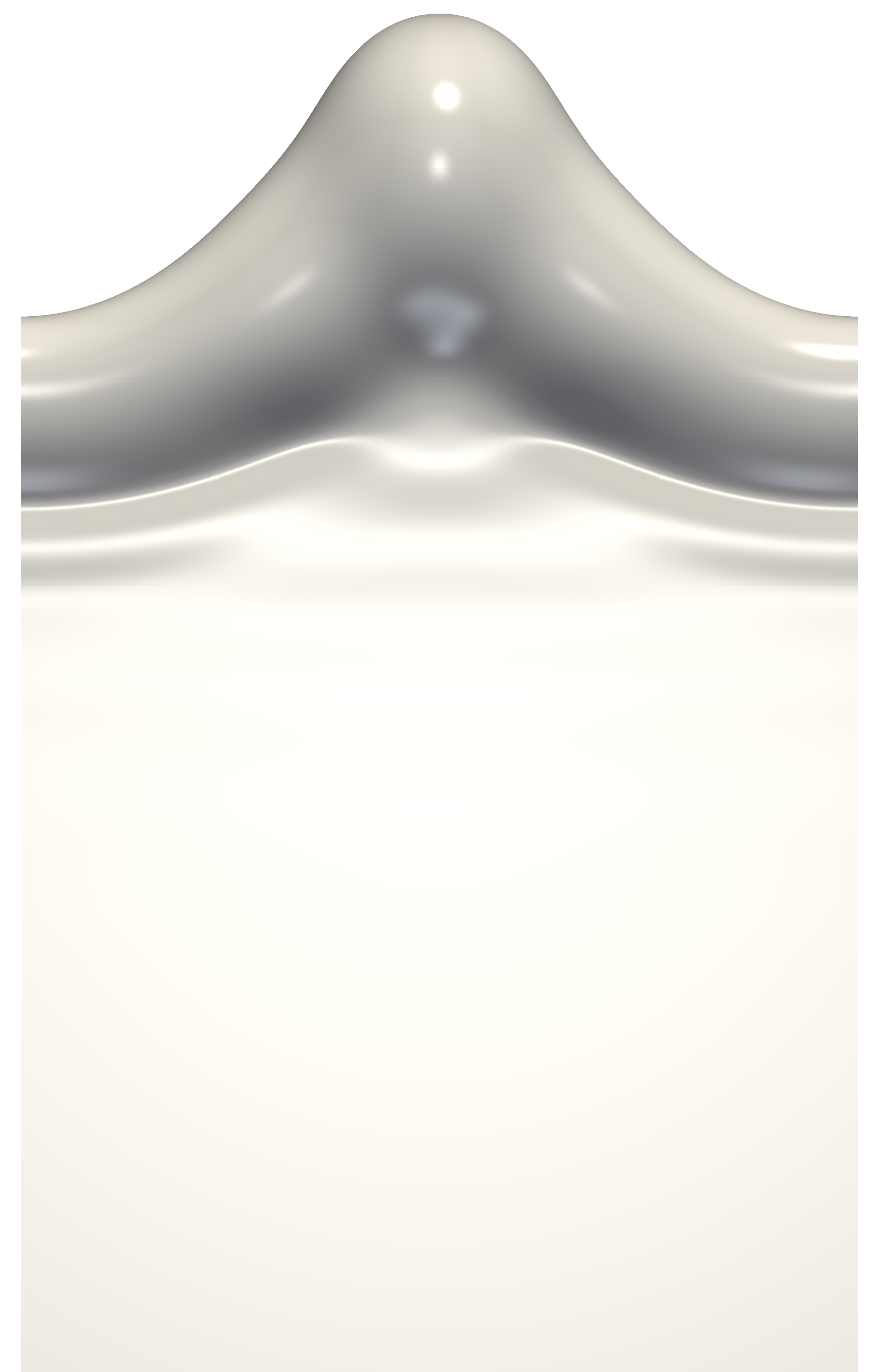}   & 
\includegraphics[width=0.15\linewidth]{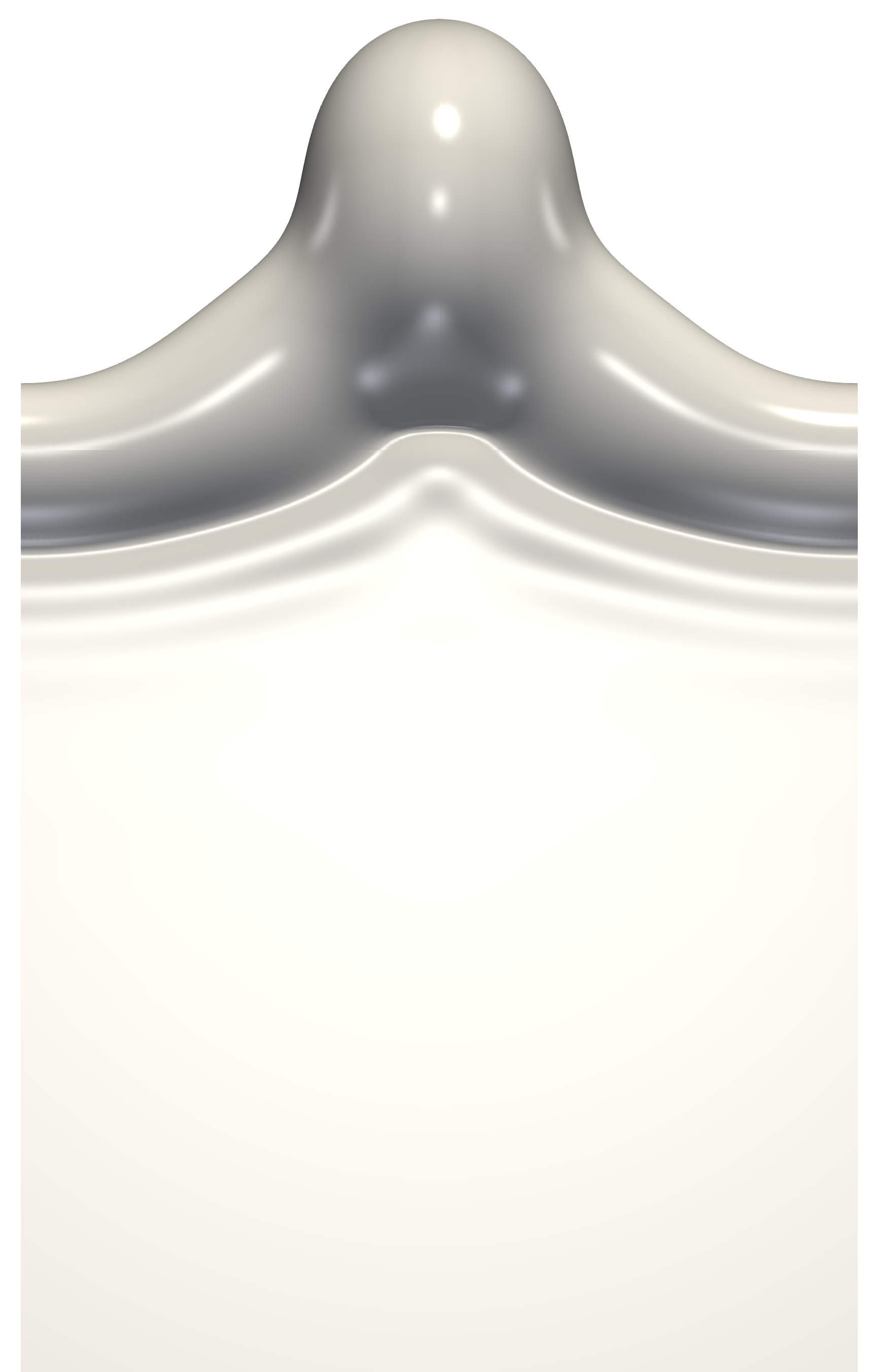}   & 
\includegraphics[width=0.15\linewidth]{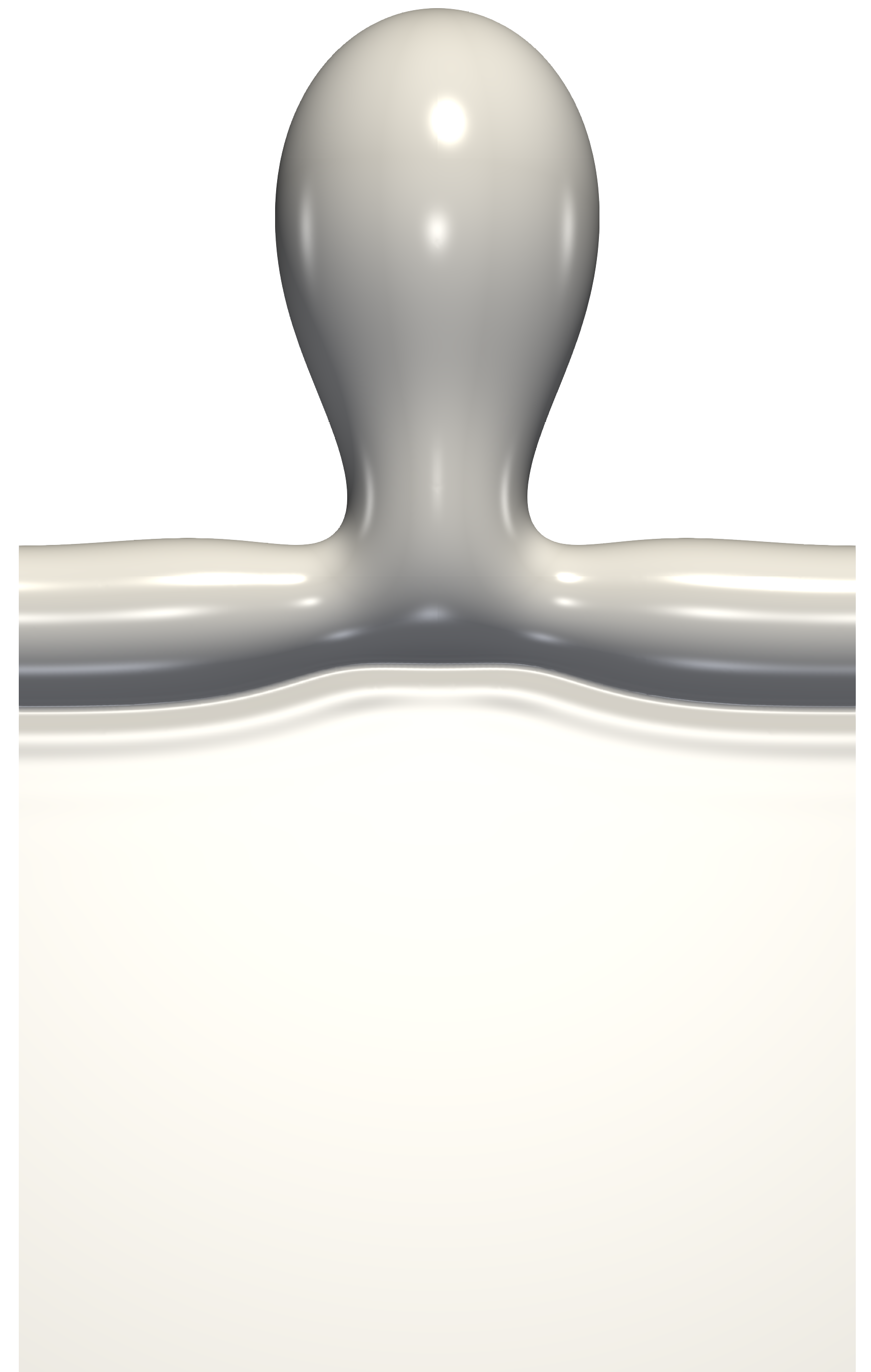}   &  \includegraphics[width=0.15\linewidth]{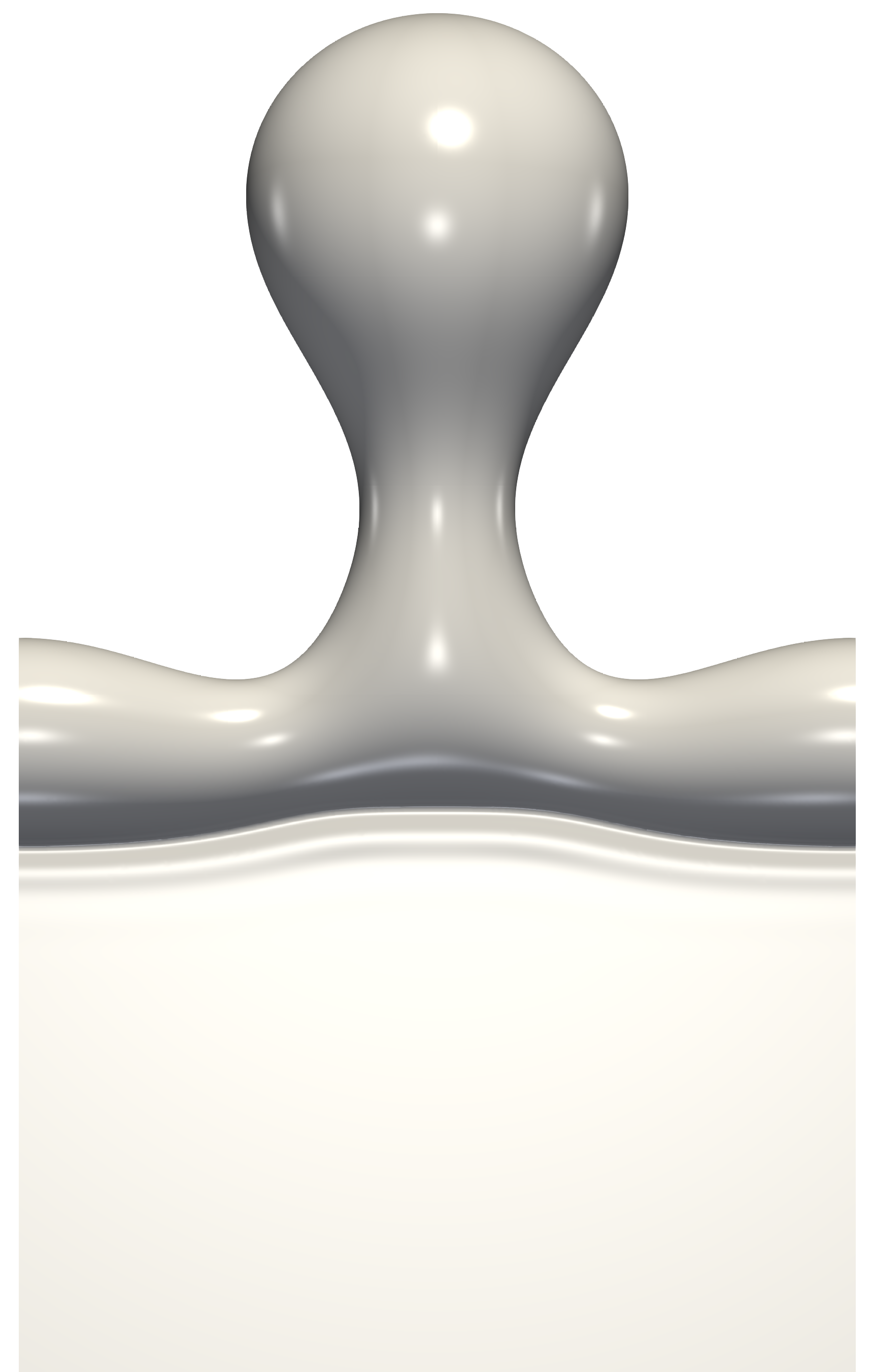}   &
\includegraphics[width=0.15\linewidth]{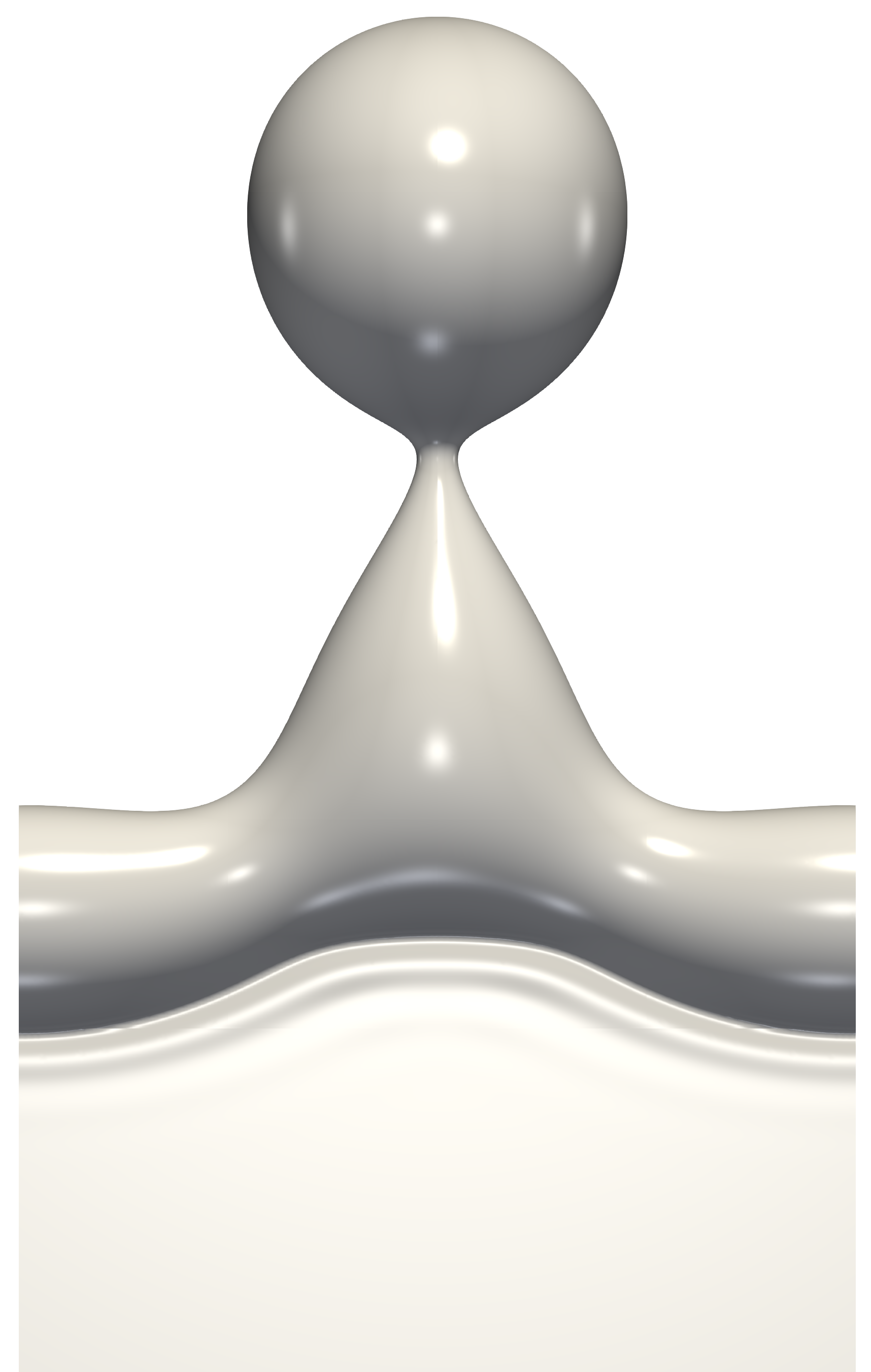}   \\
     (a) & (b) & (c) & (d) & (e) \\
 \hline
  $\beta_s=0.1$ & & & & \\
 \includegraphics[width=0.15\linewidth]{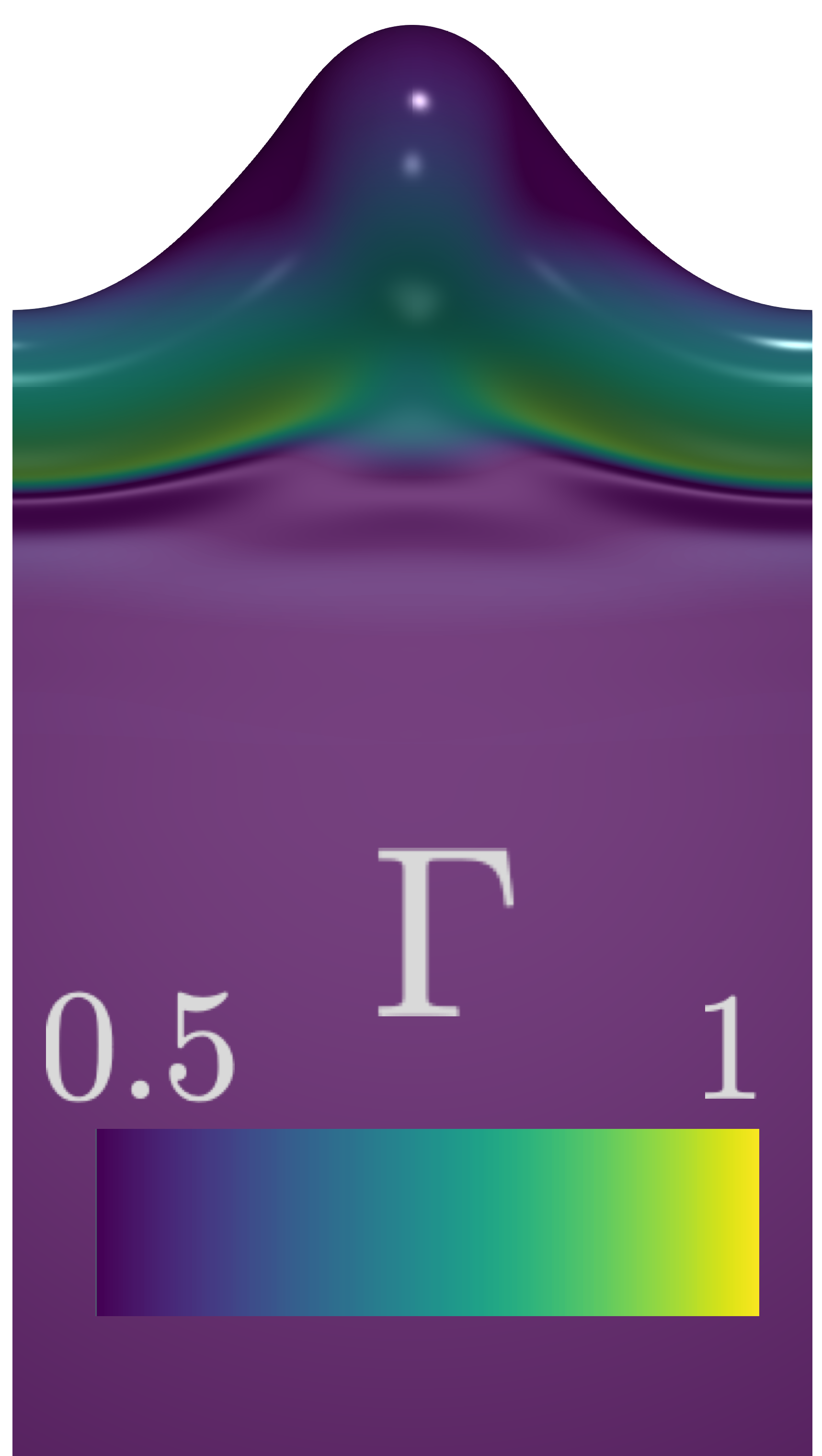}   & 
 \includegraphics[width=0.15\linewidth]{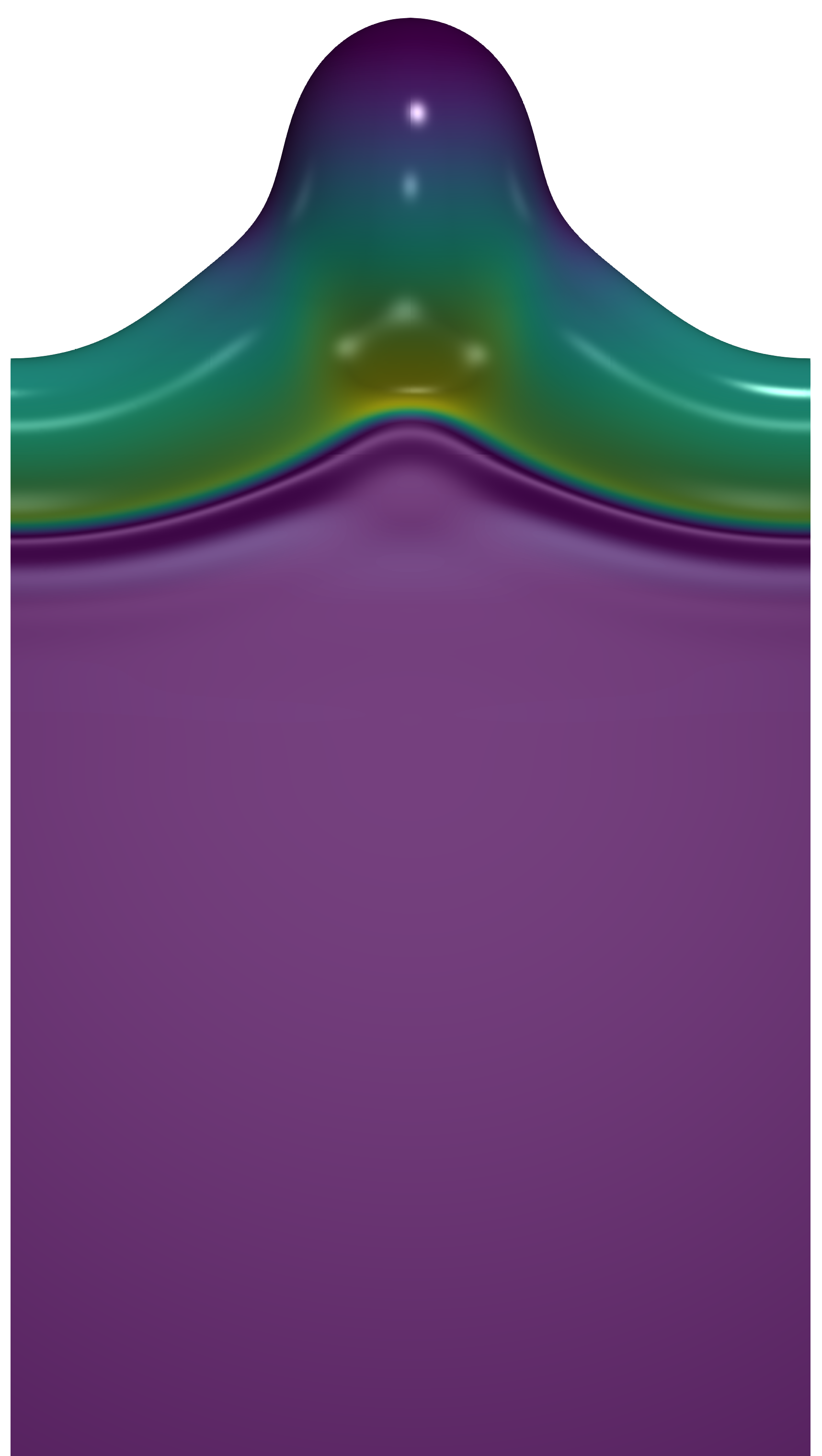}   & 
 \includegraphics[width=0.15\linewidth]{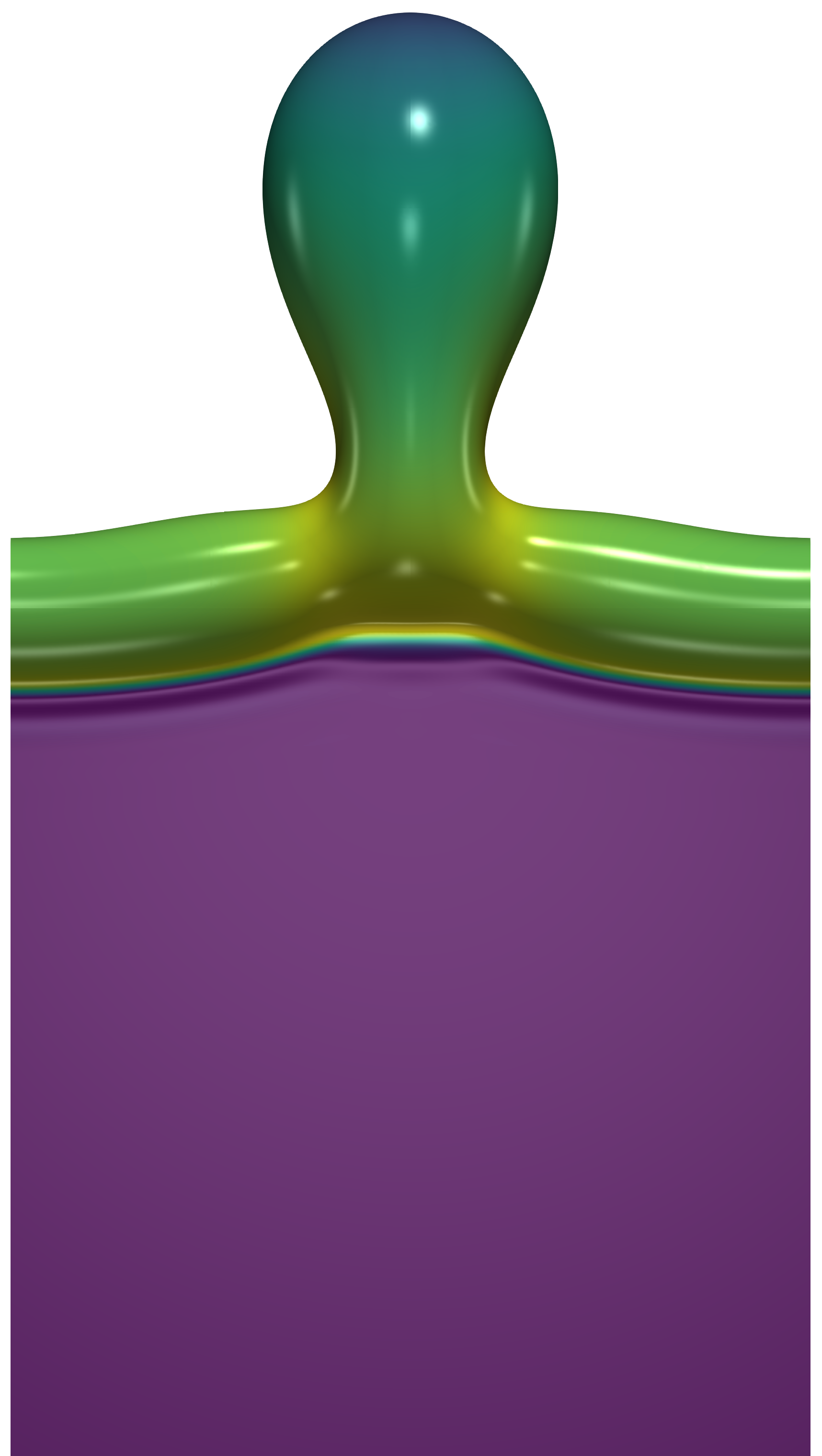}   &  
 \includegraphics[width=0.15\linewidth]{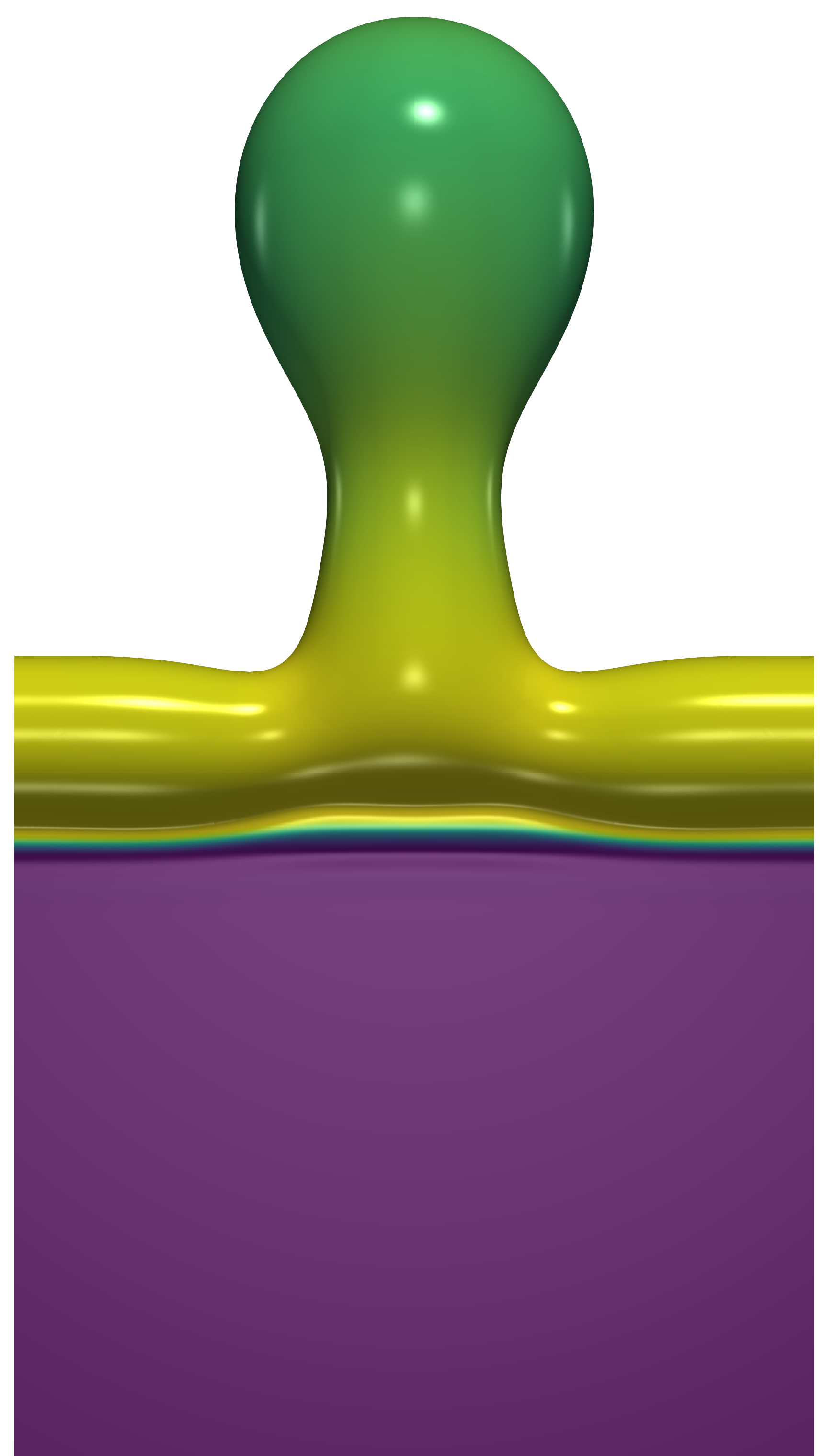}   &
 \includegraphics[width=0.15\linewidth]{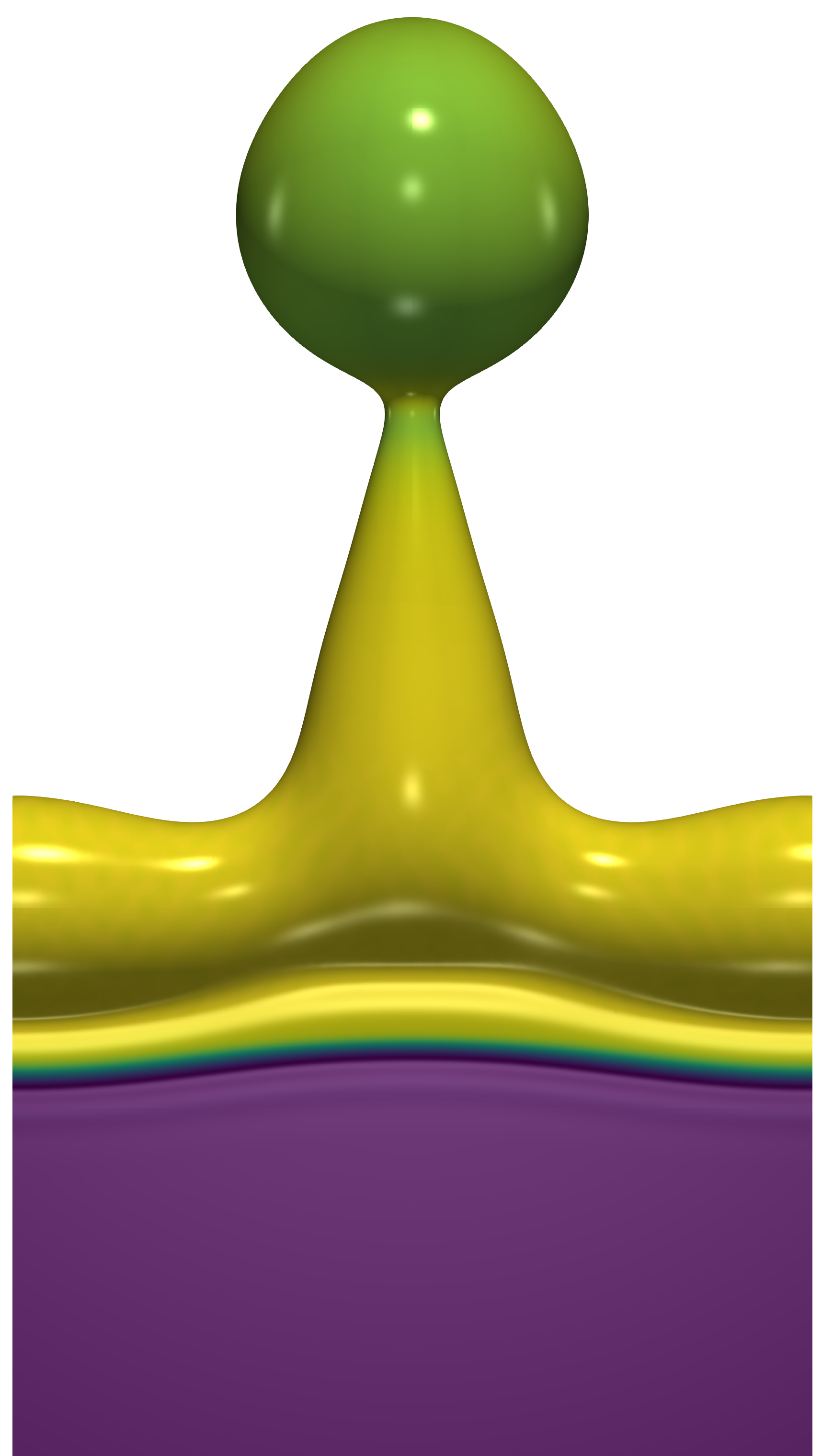}   \\
  (f) & (g) & (h) & (i) & (j) \\
  \hline
   $\beta_s=0.3$ & & & & \\
  \includegraphics[width=0.15\linewidth]{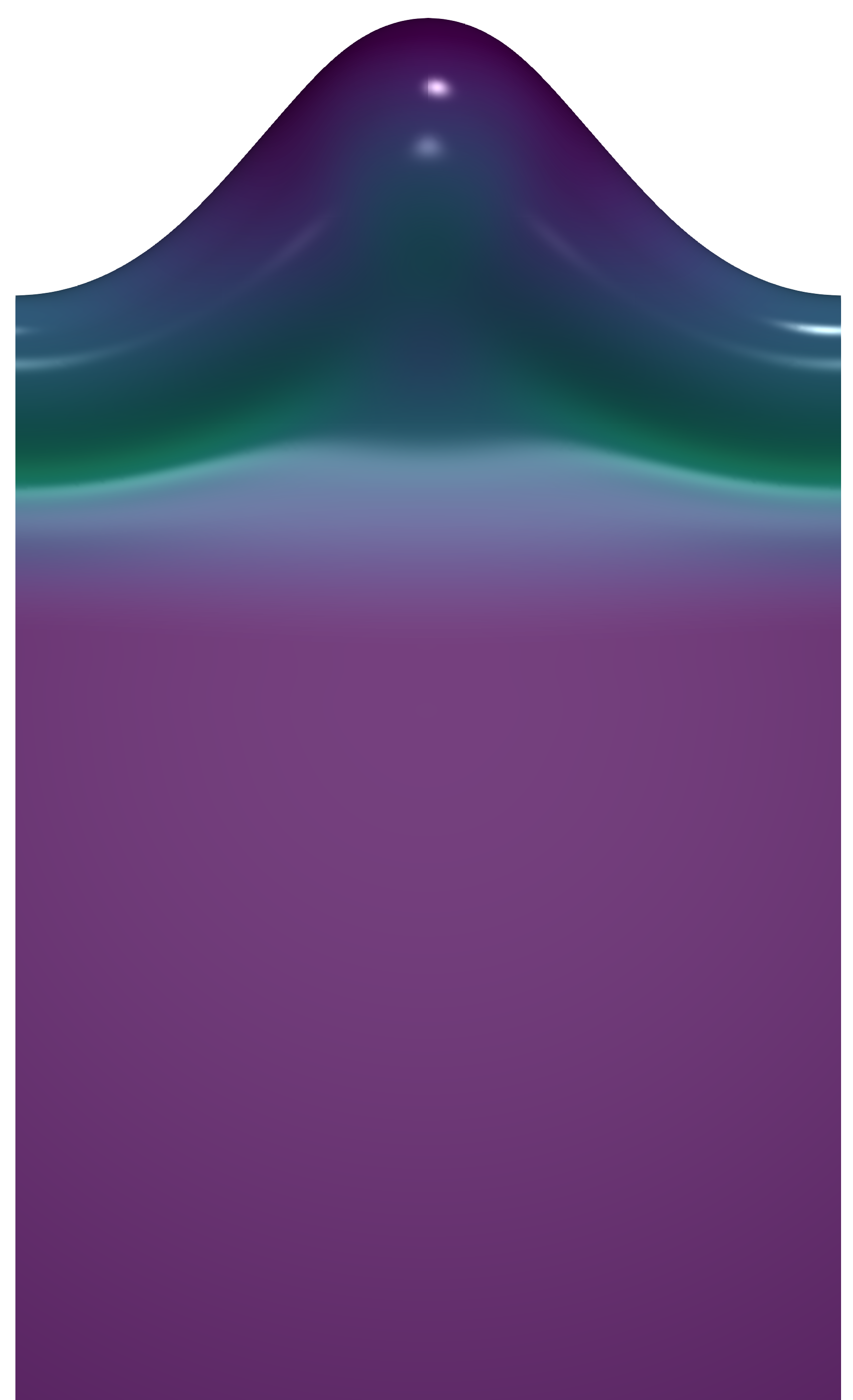}   & 
 \includegraphics[width=0.15\linewidth]{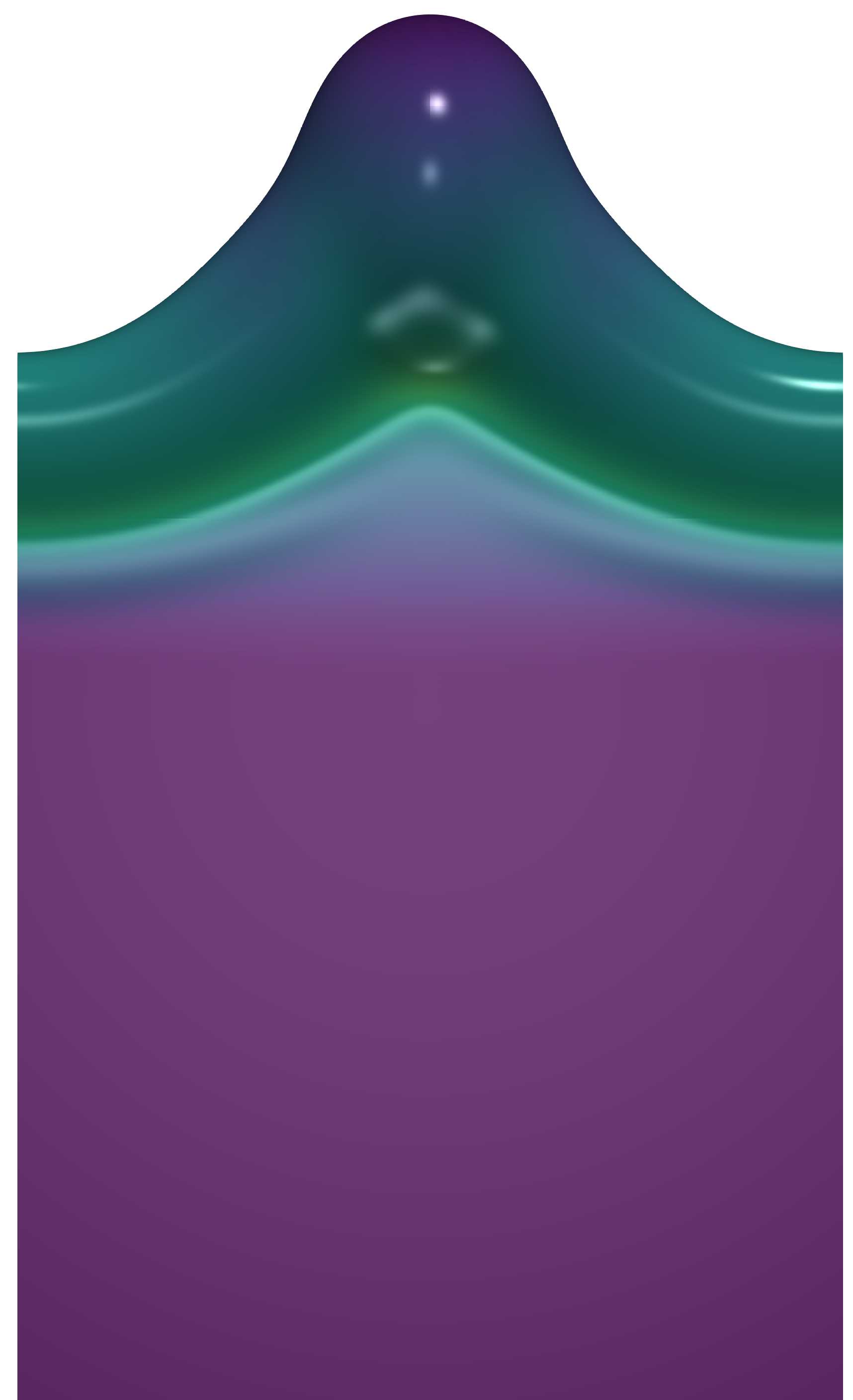}   & 
 \includegraphics[width=0.15\linewidth]{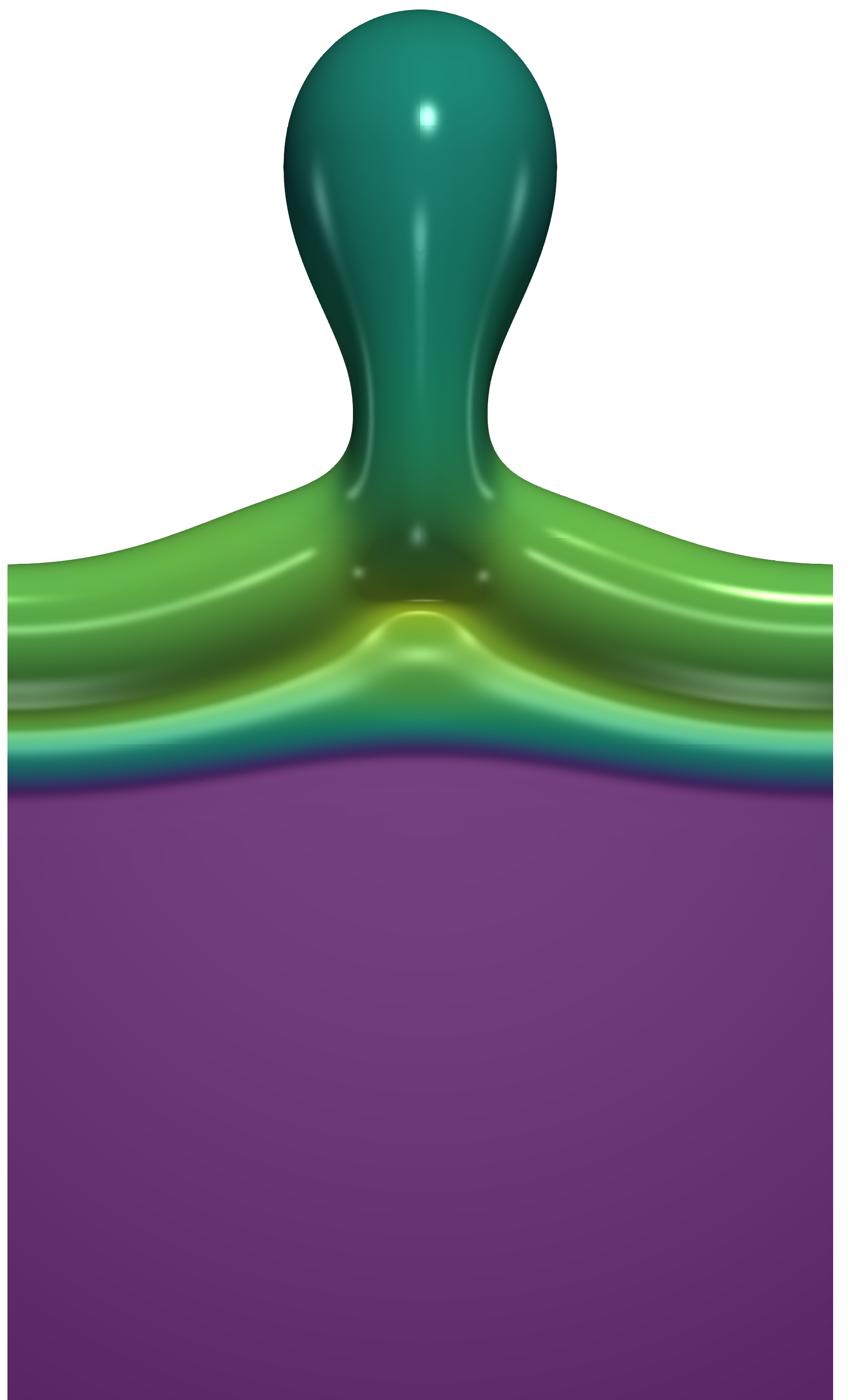}   &  
 \includegraphics[width=0.15\linewidth]{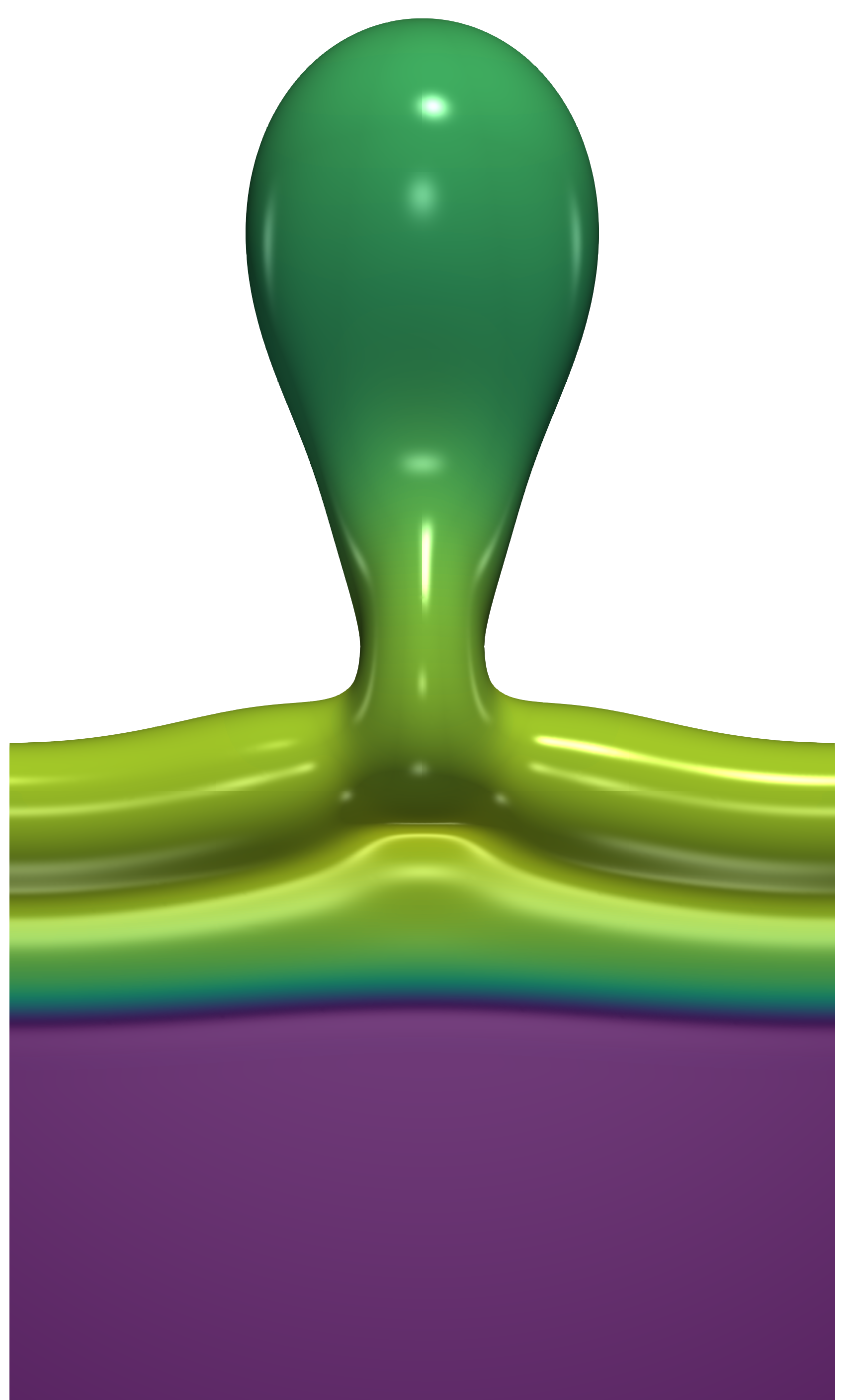}   &
 \includegraphics[width=0.15\linewidth]{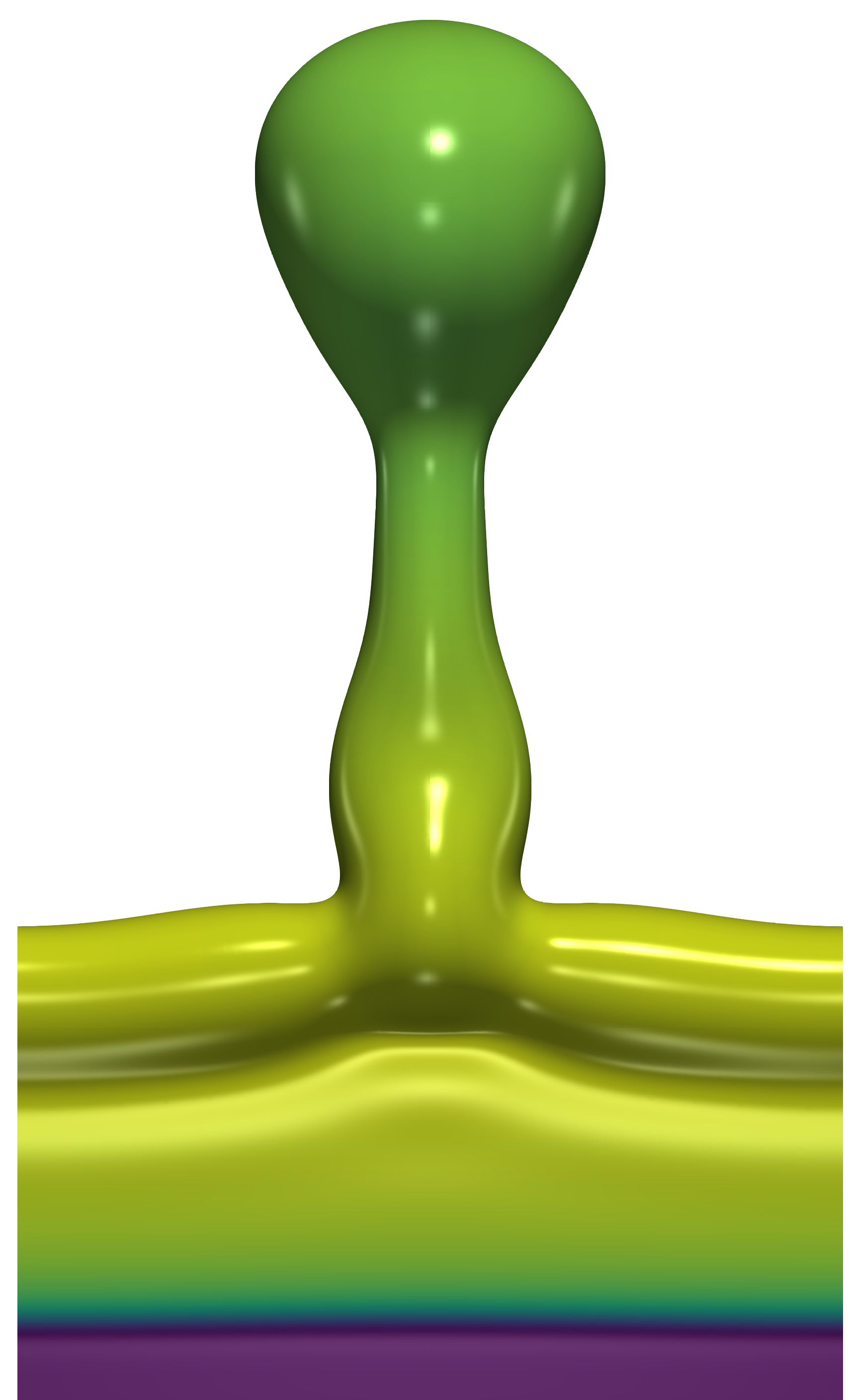}   \\
  (k) & (l) & (m) & (n) & (o) \\
  \hline
  $\beta_s=0.5$ & & & & \\
   \includegraphics[width=0.15\linewidth]{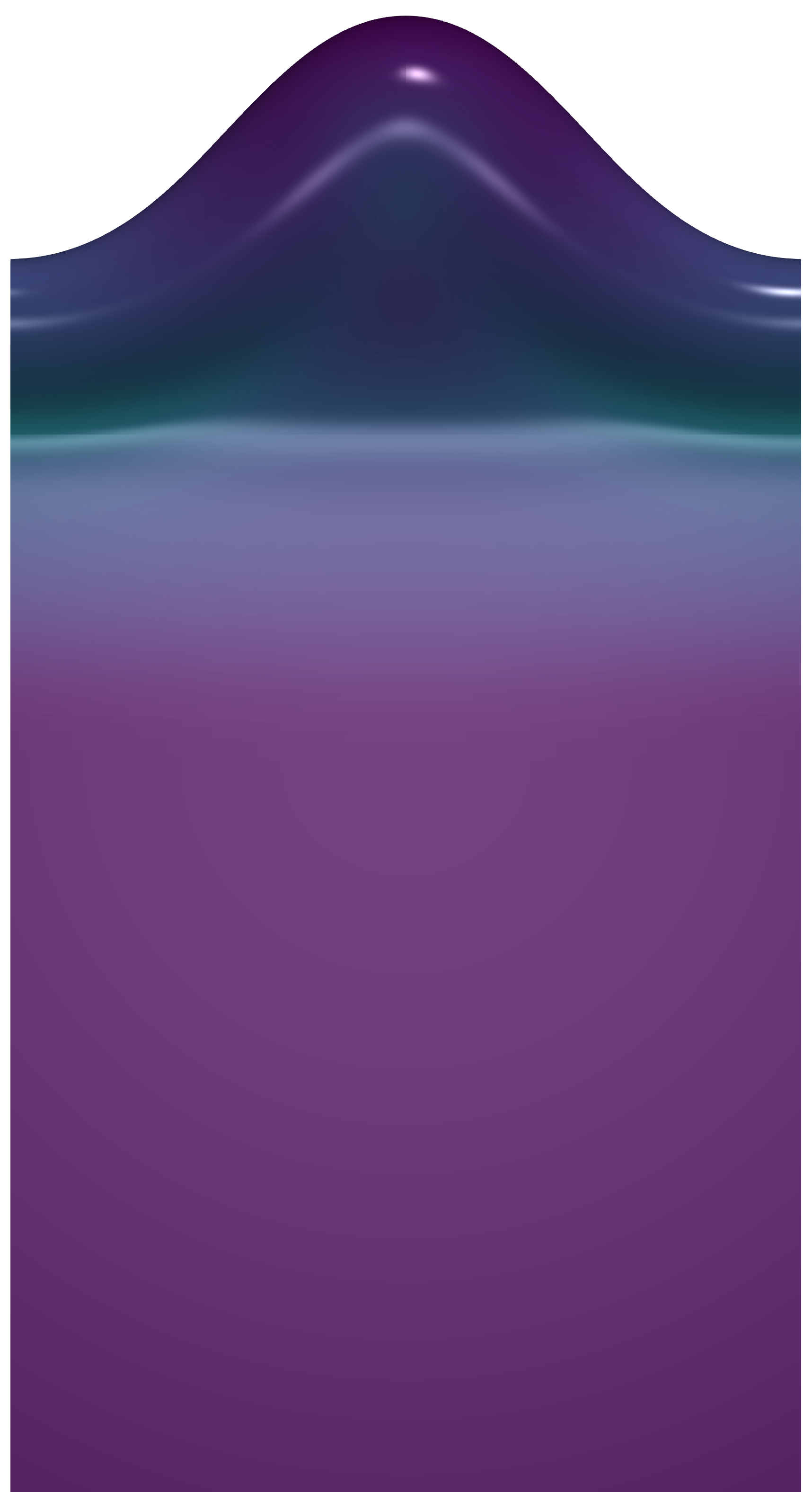}   & 
 \includegraphics[width=0.15\linewidth]{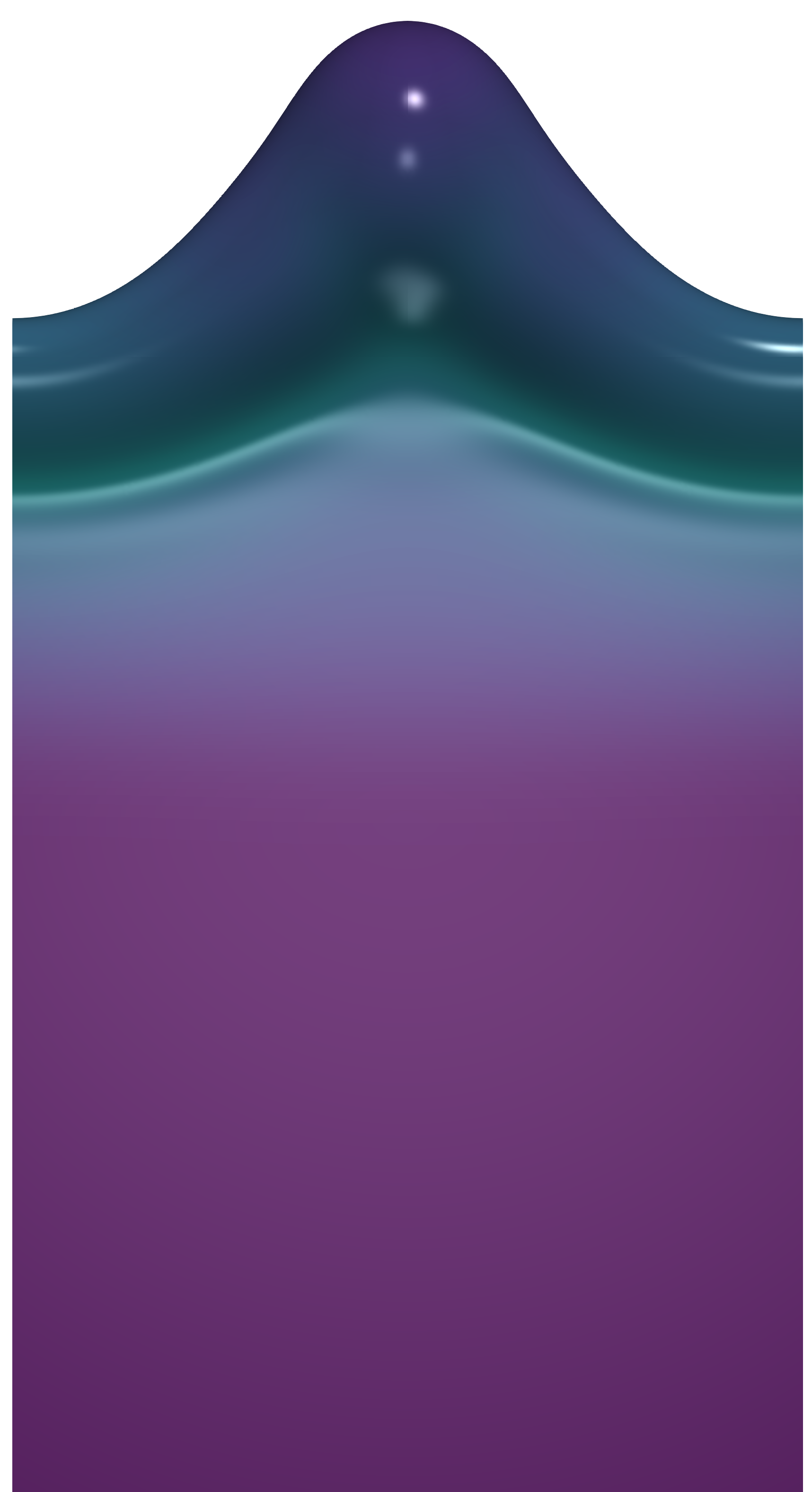}   & 
 \includegraphics[width=0.15\linewidth]{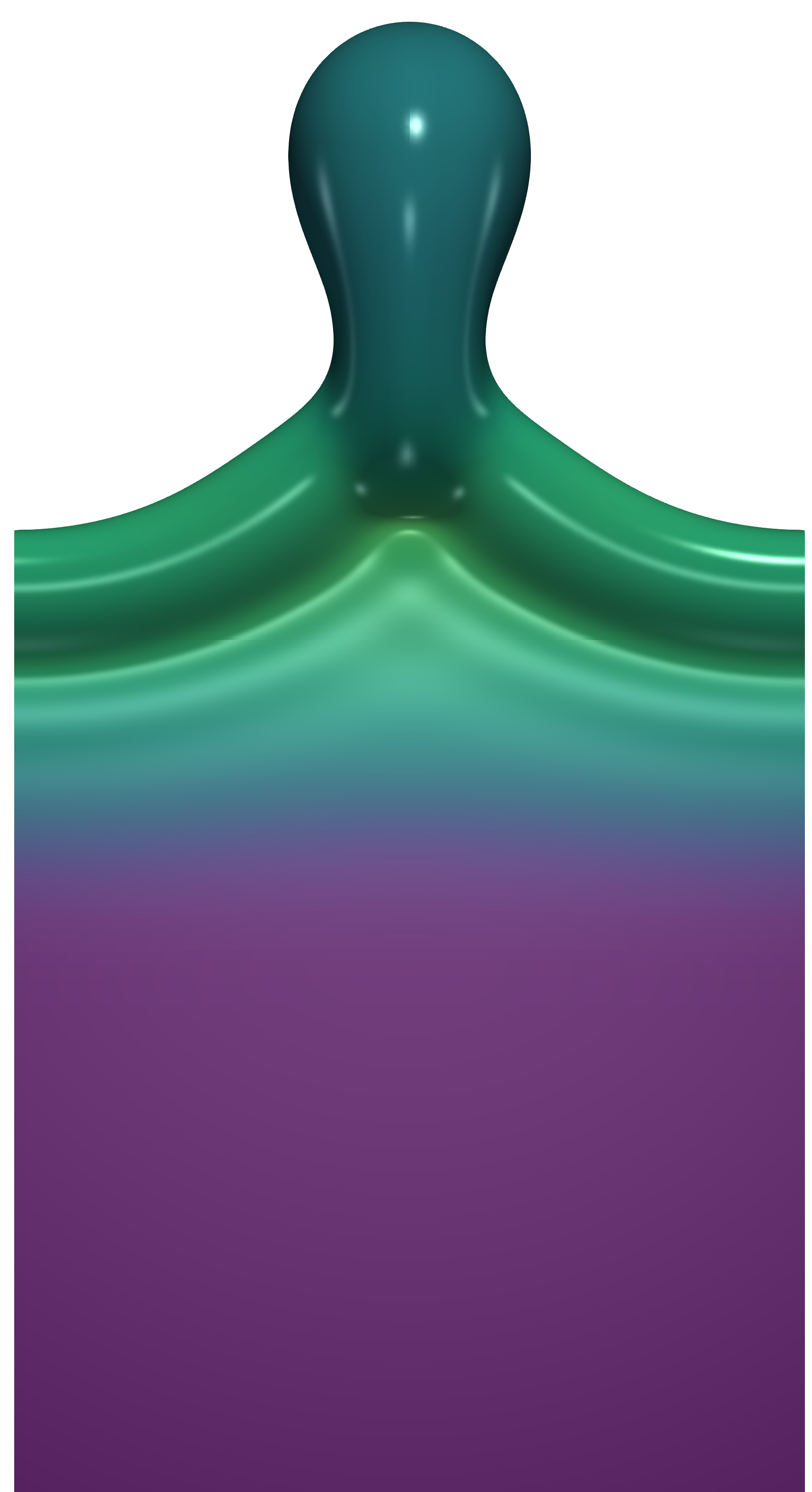}   &  
 \includegraphics[width=0.15\linewidth]{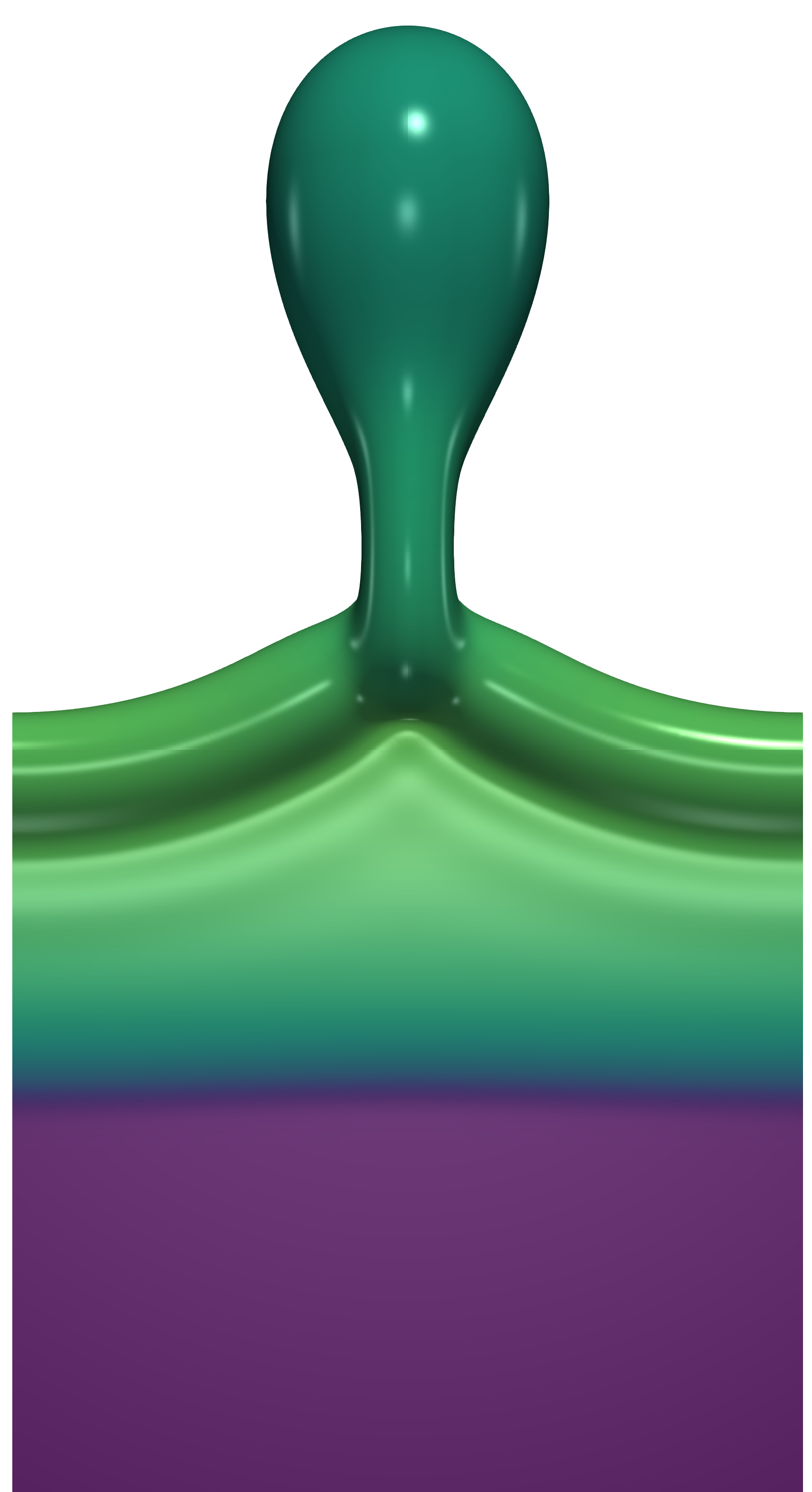}   &
 \includegraphics[width=0.15\linewidth]{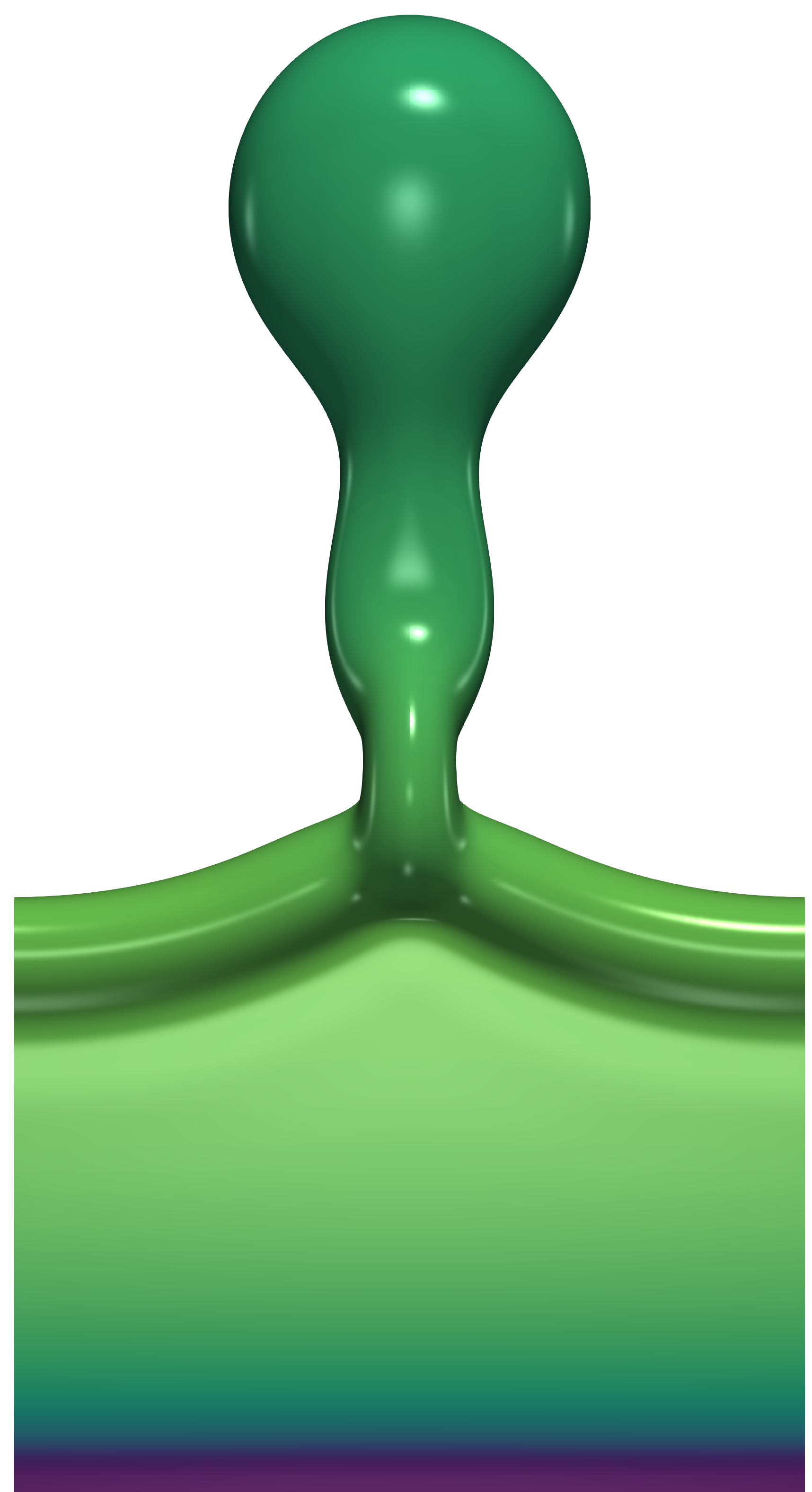}   \\
  (p) & (q) & (r) & (s) & (t) \\
\end{tabular}
\end{center} 
\caption{\label{temporal_evolution} 
Effect of surfactant elasticity ($\beta_s$) on the sheet retraction dynamics for insoluble surfactants. Here, the dimensionless parameters are $Oh = 0.0833$, $e=0.2$ and $\epsilon=0.25$; and for the surfactant-laden cases, $Pe_s=100$ and $\Gamma=\Gamma_\infty/2$. For the surfactant-laden cases, the colour indicates the surfactant concentration $\Gamma$. 
} 
\end{figure}

\begin{figure}
\begin{center} 
\begin{tabular}{cccc}
\includegraphics[width=0.25\linewidth]{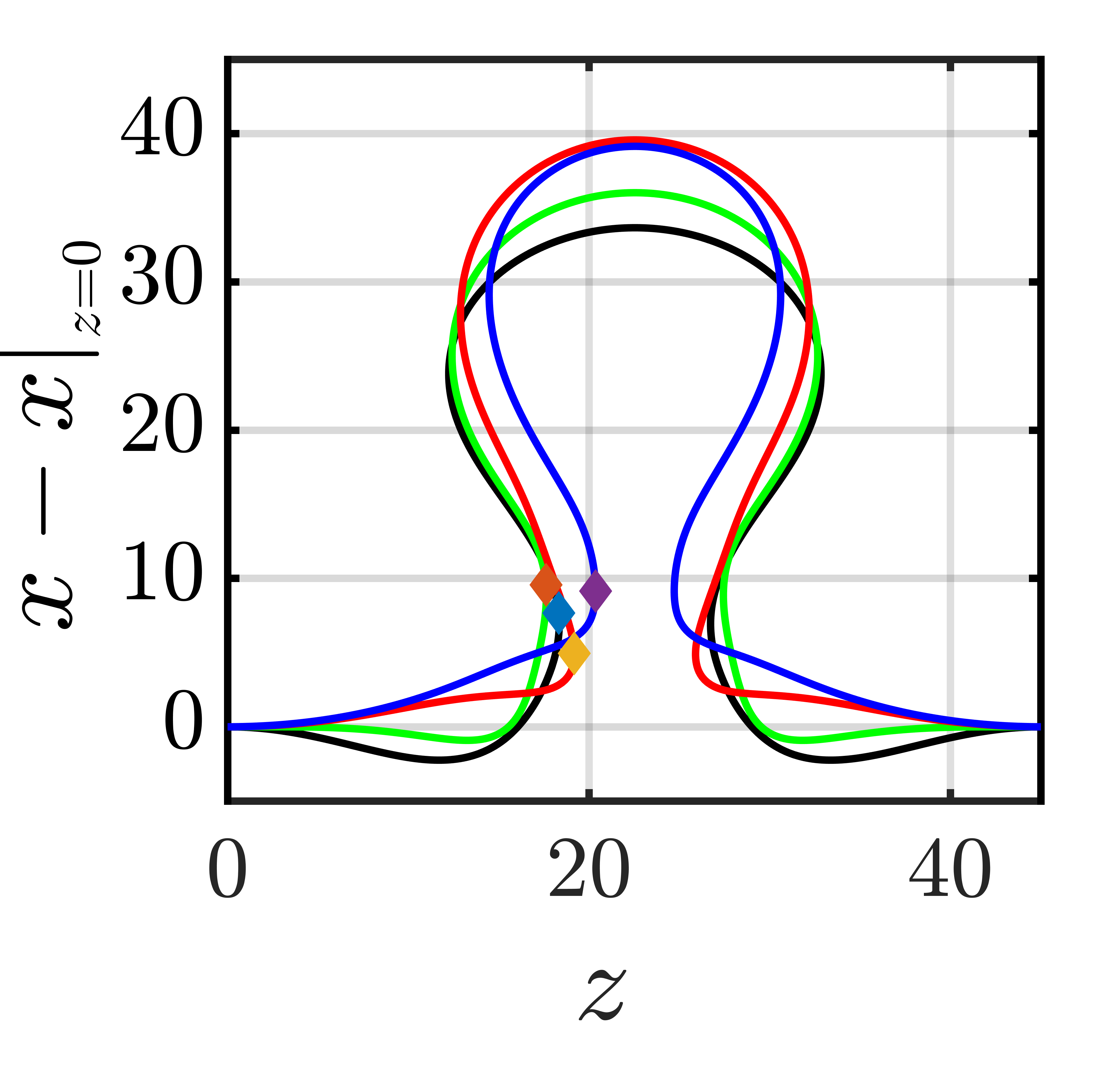}   & 
\includegraphics[width=0.25\linewidth]{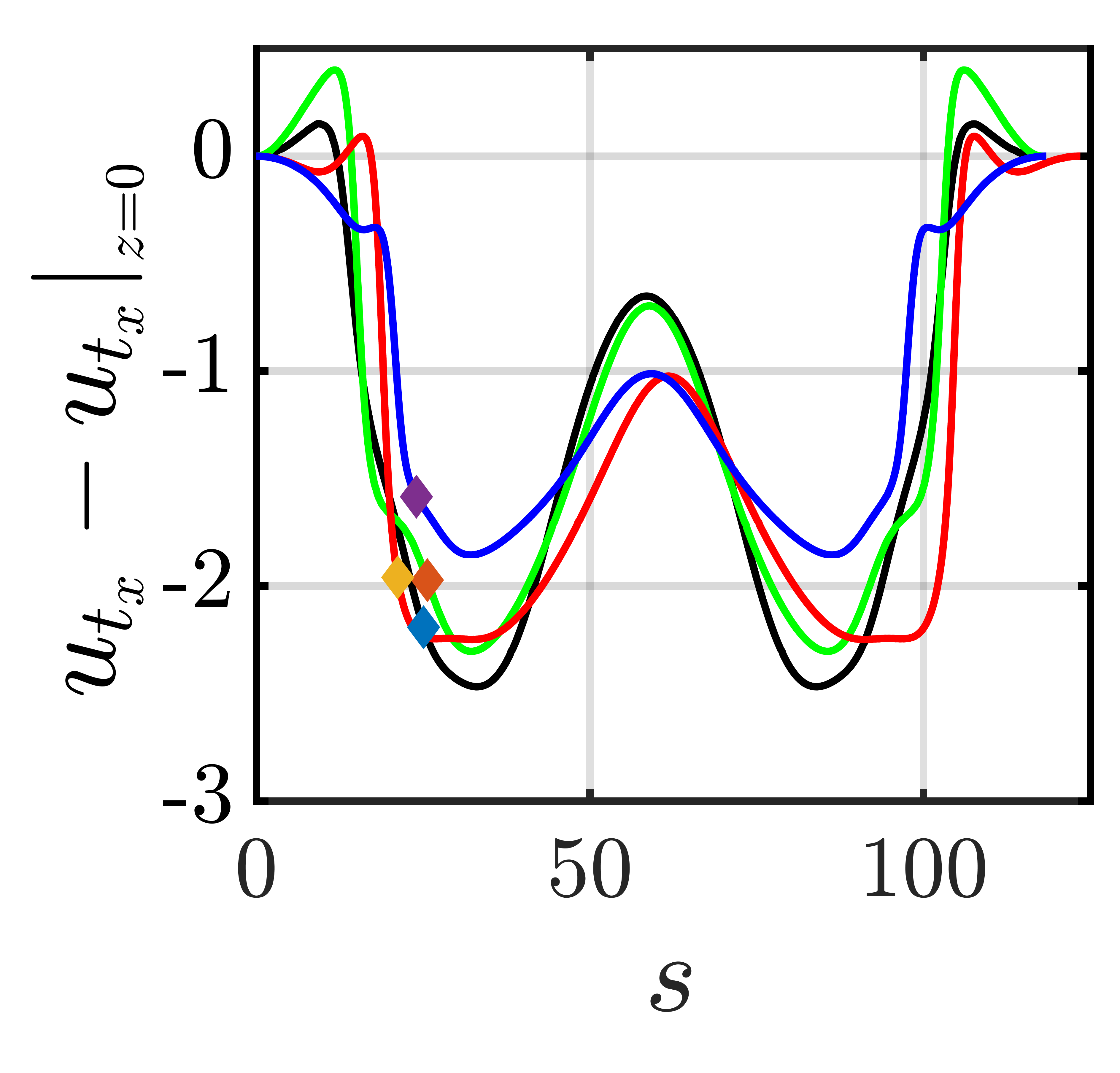}   &  \includegraphics[width=0.25\linewidth]{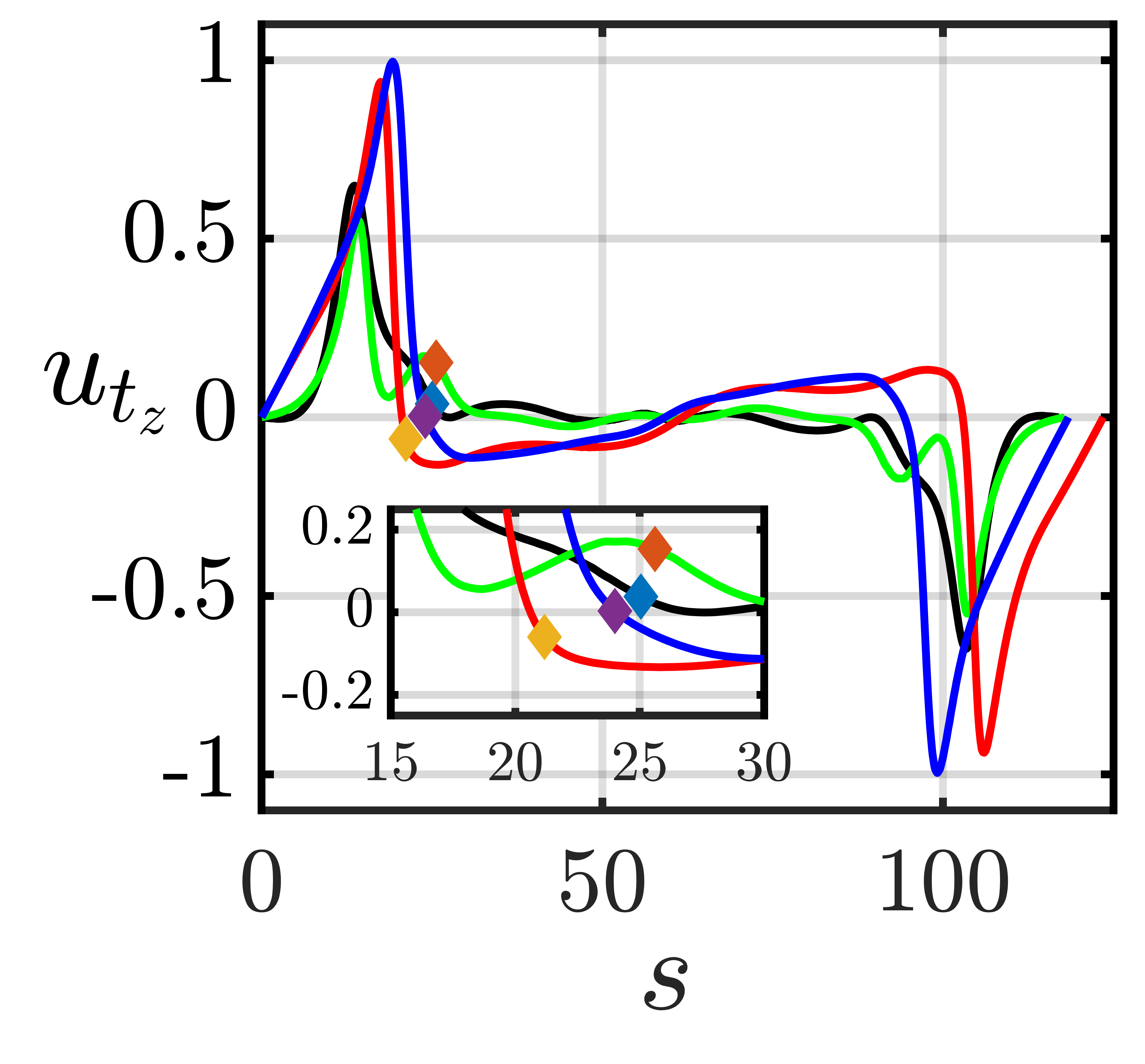}   &
\includegraphics[width=0.25\linewidth]{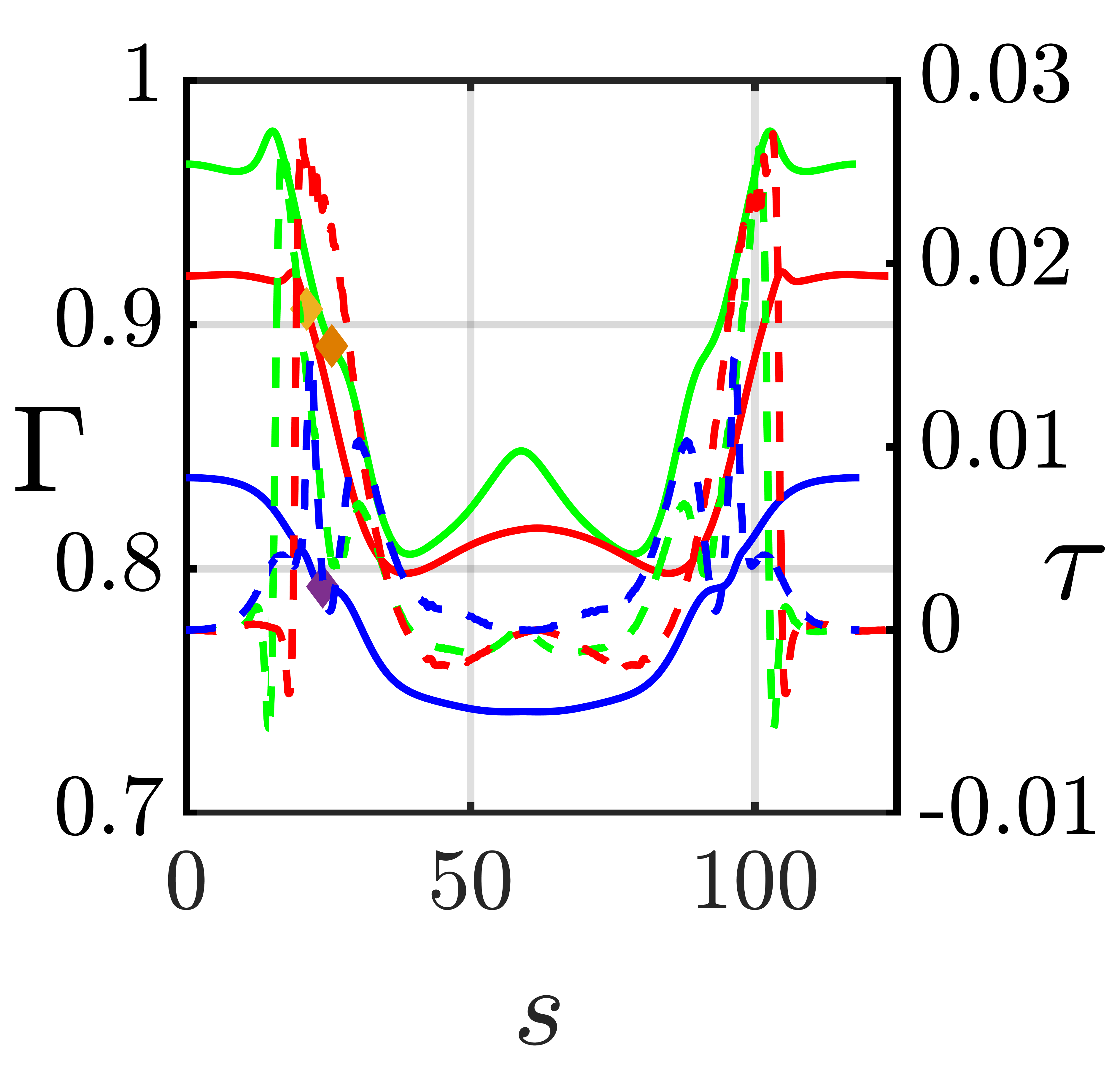}   \\
 (a) & (b) & (c) & (d)  \\
 \includegraphics[width=0.25\linewidth]{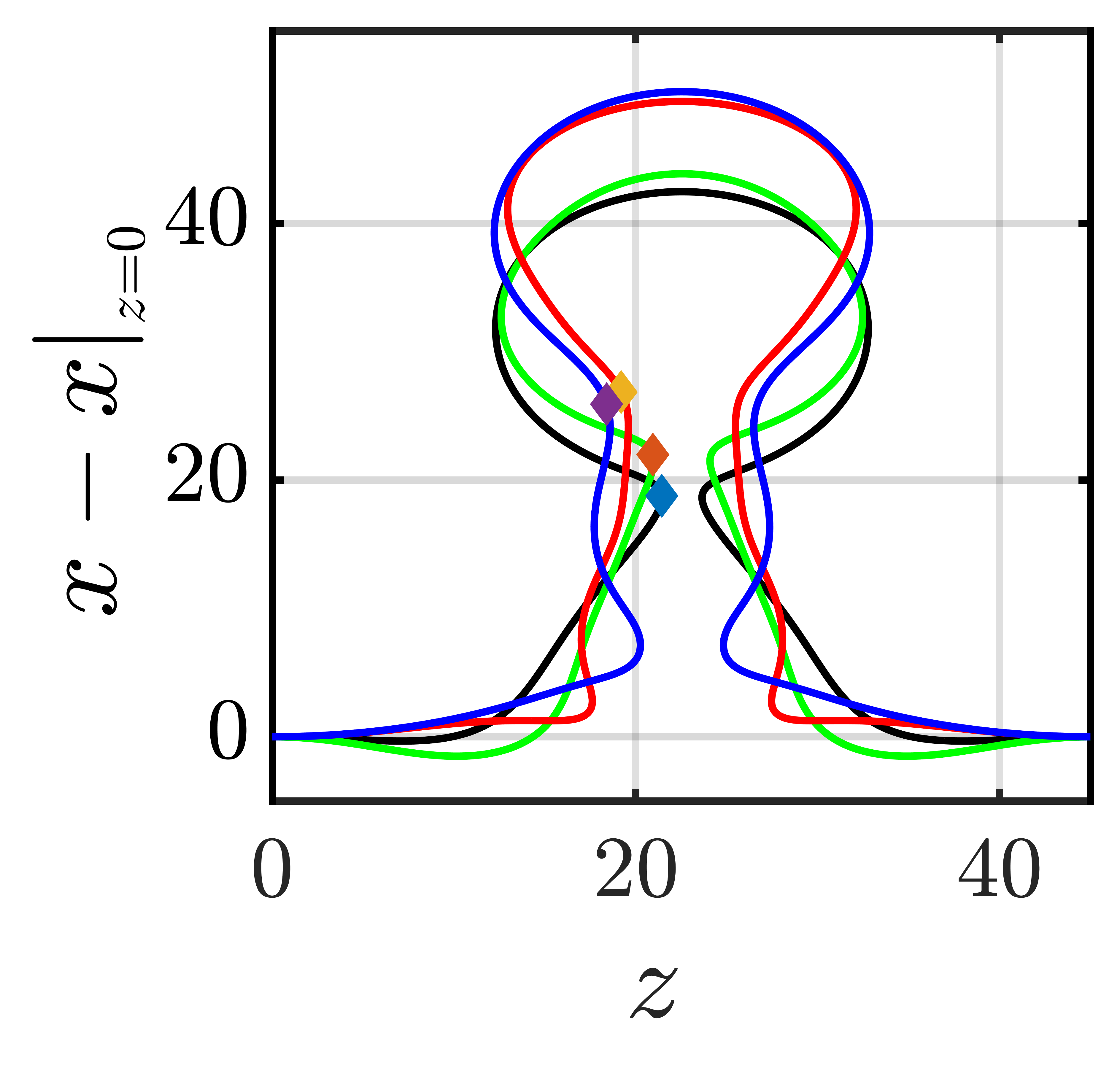}   & 
\includegraphics[width=0.25\linewidth]{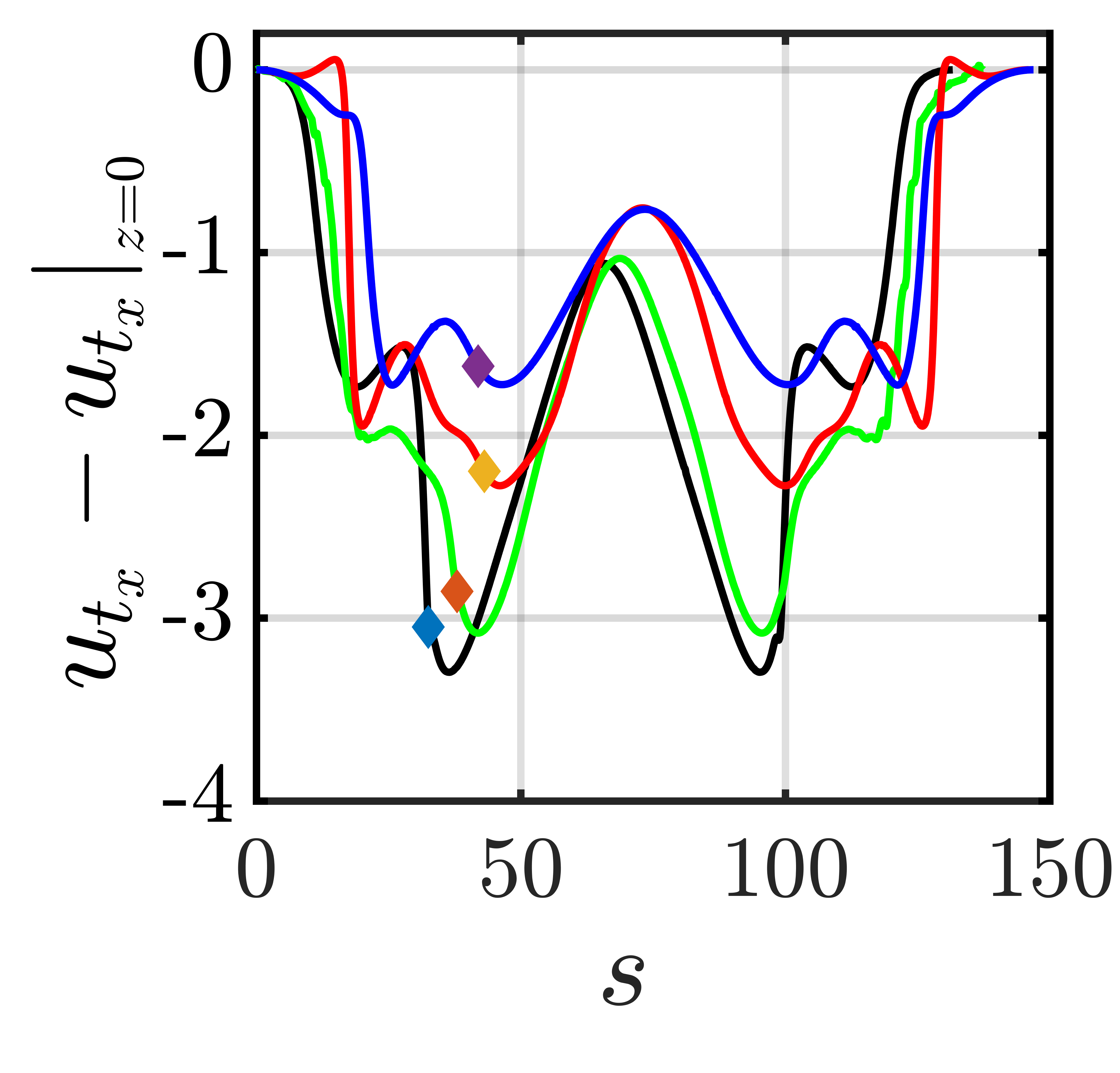}   &  \includegraphics[width=0.25\linewidth]{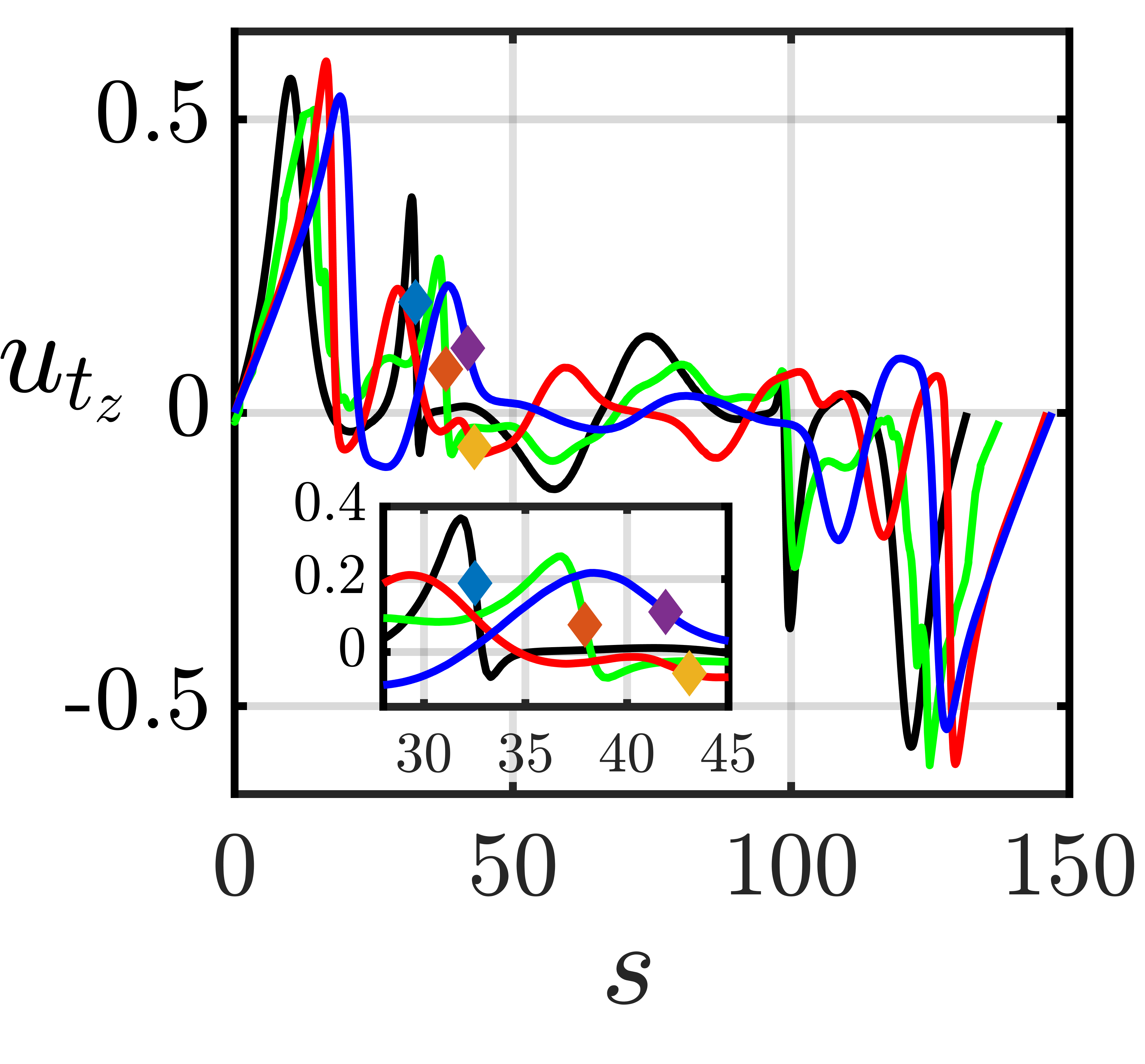}   &
\includegraphics[width=0.25\linewidth]{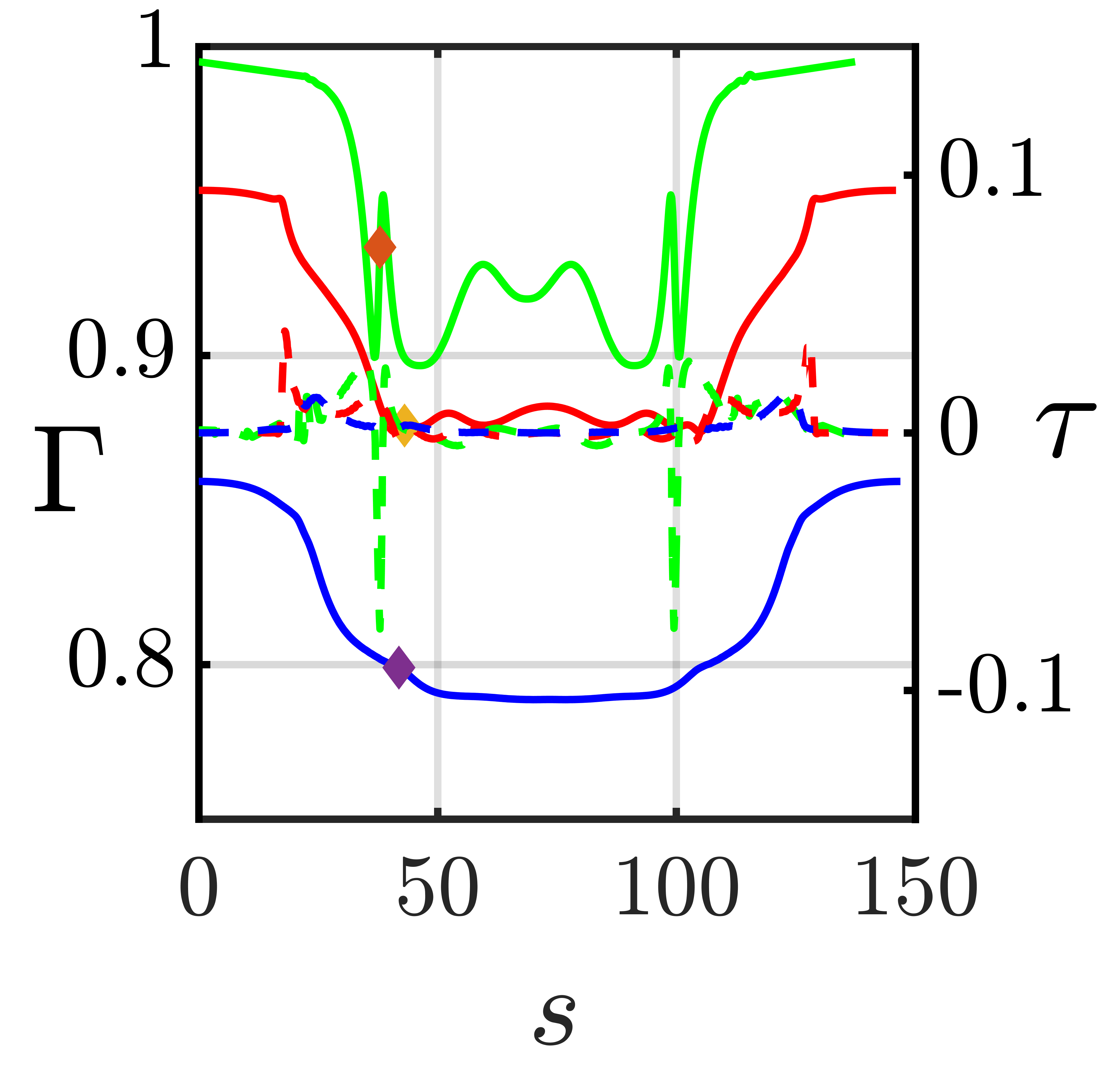}   \\
(e) & (f) & (g) & (h)  \\
\end{tabular}
\begin{tabular}{ccc}
\includegraphics[width=0.36\linewidth]{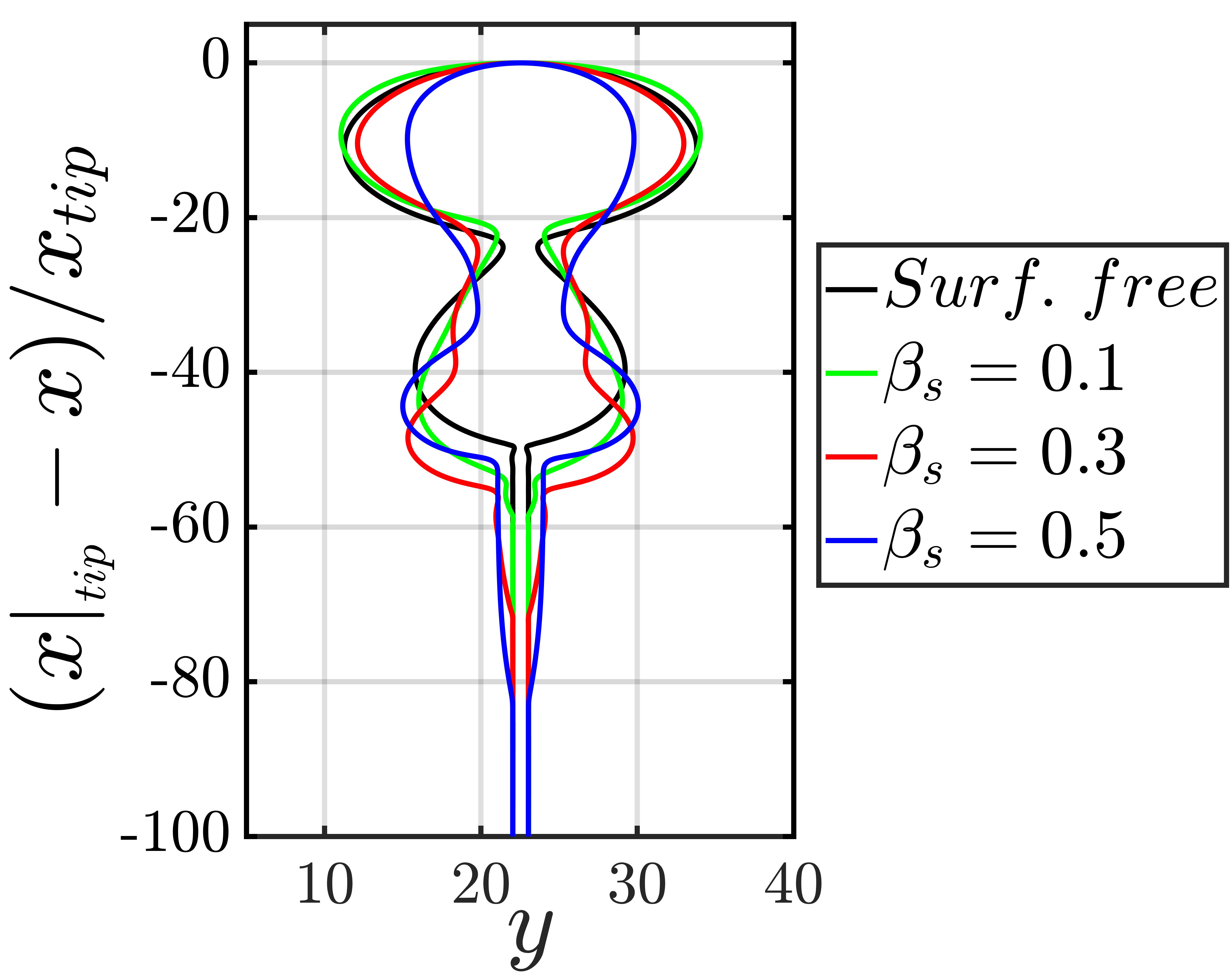}   & 
\includegraphics[width=0.17\linewidth]{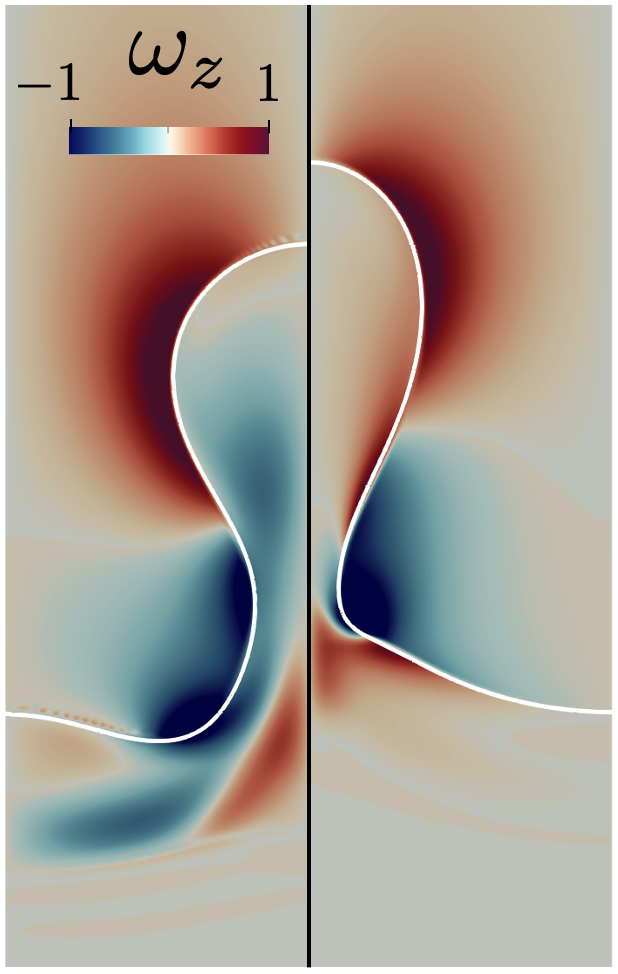}   & 
\includegraphics[width=0.17\linewidth]{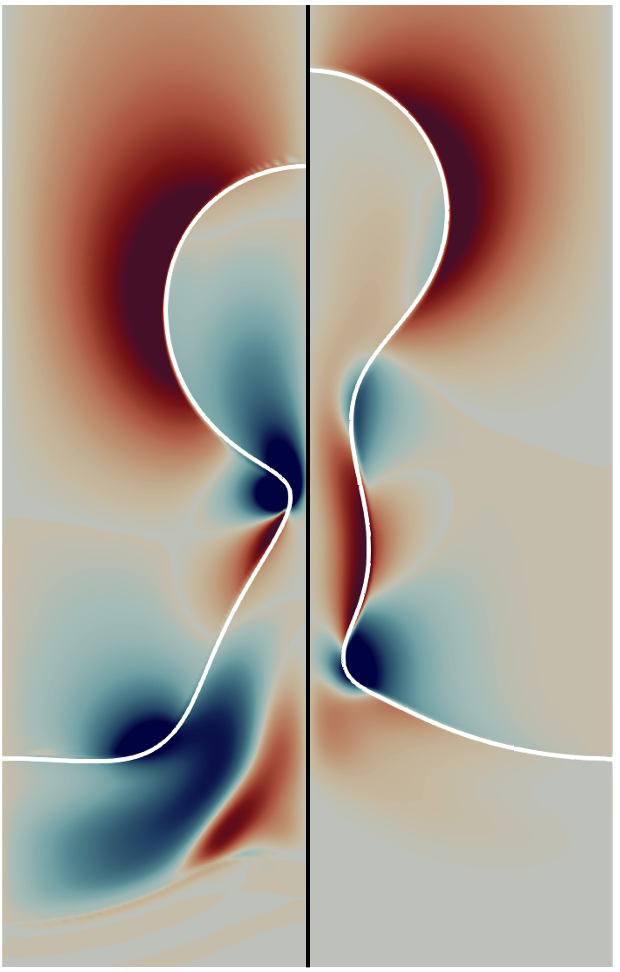}   \\ 
 (i) & (j) & (k) \\
\end{tabular}
\end{center} 
\caption{\label{s_plot} 
Panels $(a-h)$: Effect of the elasticity parameter, $\beta_s$, on the flow and  surfactant concentration during sheet retraction. 
Two-dimensional projection of the interface, $\Gamma$ and $\tau$, $u_{tx}$ and $u_{tz}$ in the $x–z$ plane ($y=\lambda/2$) are shown in $(a–d)$ and $(e–h)$ for $t=141.42$ and  $t=189.50$, respectively.
Note that the abscissa in (a,e) corresponds to the $z$ coordinate,
and in $(b–d)$ and $(f–h)$ to the arclength, $s$. The diamond shapes  show the location of the bulbous necks.
Panel (i) shows the effect of $\beta_s$ on the film thickness of the sheet in the $x–y$ plane ($z=\lambda/2$).
Panels $(j-k)$ show the effect of surfactants on the vorticity, $\omega_z$,  for the surfactant-free case (left panels), and the surfactant-laden case (right panels), for $\beta_s=0.5$ at  $t=141.42$ and  $t=189.50$, respectively.
} 
\end{figure}


The spatio-temporal interfacial dynamics for the surfactant-free case,  with $Oh=0.0833$, are found in Figures \ref{temporal_evolution}a-e.
As seen, the initial rim instability grows to develop nonlinearities that eventually lead to a capillary singularity (break up).
At early stages of the simulation, the fluid is pulled from the rim  to form an elongated ligament (see figure \ref{temporal_evolution}c).
Over time, the ligament undergoes 
end-pinching, where capillarity overcomes viscosity to form a bulbous end (see figure \ref{temporal_evolution}d) terminating in the break up of a droplet (see figure \ref{temporal_evolution}e). The interfacial shape prior to the capillary singularity closely resembles experimental observations by \citet{Zhang_2010} and \citet{wang_bourouiba_2018,wang_bourouiba_2021}.
Additionally, we observe surface waves resulting from  the development of capillary instabilities at the upstream of the rim. 

Attention is now turned to the effect of surface-active agents by varying the surfactant elasticity parameter,  $\beta_s$, with $Oh=0.0833$, $Pe_s=100$ and $\Gamma=\Gamma_\infty/2$.
For $\beta_s=0.1$, the spatio-temporal interfacial dynamics resemble that of a surfactant-free case, from the formation of the ligament to the eventual end-pinching: the dynamics results in droplet shedding (see figure \ref{temporal_evolution}f-j). 
In this case, surfactant is accumulated at the base of the cylindrical rim as it gradually increases its size. Once the ligament is formed, $\tau$ induces a flow towards the ligament (see figure \ref{temporal_evolution}h) resulting in a longer ligament prior to pinch-off. However, as surfactant-induced Marangoni-stresses are not sufficiently strong to evacuate $\Gamma$, the dynamics results in modest surfactant concentration gradients, and,
in agreement with \citet{Constante-Amores_prf_2020}, the ligaments breaks up. 
By increasing $\beta_s$, $\tau$ act to reduce $\Gamma$-gradients by inducing a flow from the rim towards the sheet, and from the rim towards the ligament  (regions of low surfactant concentration), resulting in a severe reduction of the rim radius (displayed in figure \ref{temporal_evolution}).
The surfactant-driven  flow  leads to the
reopening of the fluid sheet adjacent to the retracting rim thereby increasing its film thickness (see figure \ref{s_plot}i); the Marangoni-induced flow also acts to
suppress the capillary waves ahead of the rim (see figure \ref{temporal_evolution}k-o and \ref{temporal_evolution}j-n for $\beta_s=0.3$ and $\beta_s=0.5$, respectively).

We now turn our attention to the two-dimensional interfacial shape represented by $\Gamma$ and spanwise Marangoni stresses, the tangential component of the surface velocity $u_{t_{z}}$, the streamwise surface velocity, $u_{t_{x}}$ (where the velocity is given in the reference frame of $u_x$ at $z=0$)
presented in figure \ref{s_plot} at $t=141.42$ and  $t=189.50$, respectively. 
As observed, the surfactant-driven flow, on the spanwise direction, results on the retardation of the dynamics brought by surfactant-induced interfacial rigidification (see the dampening of $u_{t_{x}}$ in the frame of difference in figure \ref{s_plot}b,f: as $\beta_s>0$, $u_{t_{x}} < 0$). 
We observe that surfactant concentration gradients at $t=141.42$ give rise to Marangoni stresses which drive a flow from the horizontal part of the rim towards the base of the ligament, and from the tip of the ligament to its base (for high $\beta_s$, the ligament tip has a nearly uniform distribution of $\Gamma$). 
The combined effect of Marangoni stresses is to bridge the surfactant gradient between the tip of the ligament and the rim (see  figure \ref{s_plot}h).
This effect is sufficiently strong to promote the development of a thinner bulbous end, and cause a
surfactant-driven escape from end-pinching to a longer ligament (in agreement with \citet{Constante-Amores_prf_2020} and \citet{kamat_2020}). As expected, in a latter stage, the nearly uniform distribution of surfactant concentration results in the elimination of Marangoni stresses (see figure  \ref{s_plot}h). 
Figures \ref{s_plot}c and g show that that $u_{t_{z}}> 0$ upstream and $u_{t_{z}}  < 0$ downstream of the neck, i.e. an stagnation point is found in between. 
As the neck narrows, capillary-driven flow causes further neck thinning and the creation of a velocity maximum at $u_{t_{z}}$, as illustrated in figure \ref{s_plot}c,g  which is indicative of singularity formation. The $u_{t_{z}}$ profile for the surfactant-free case is characterized by the presence of a large velocity maximum and two stagnation points, 
with the neck located in between.
By close inspection of the $u_{t_{z}}$ profile of the surfactant-laden cases it becomes clear that only one stagnation point is found near the neck for high $\beta_s$. 
The surfactant-induced Marangoni stresses result in the suppression of one of the stagnation points-- in agreement with \citet{Constante-Amores_prf_2020,constanteamores2020bb}.

Figure \ref{s_plot}j,k shows the vorticity field, i.e. $\omega=\triangledown \times \textbf{u}$,  in terms of the interface location for the surfactant-free and surfactant-laden cases for $t=141.42$ and $t=189.50$ (the location is given in a reference frame where the the rim position is found at $z=0$). 
For all cases, a primary vortex (with a positive sign) is formed at the front region of the bulbous edge and is shed into the ambient fluid.
For the surfactant-free case, we observe a negative vortex-ring near the bulbous neck resulting from a change of curvature; this vortex ring grows over time ending in breakup. 
For the surfactant-laden case at $t=141.42$, $\tau$ induce vorticity from the bulbous tip to the neck and from the rim to the ligament. This flow causes the neck to reopen; after escaping the singularity the  
flow through the neck triggers the formation of a jet towards the centre of the bulbous region (these findings are consistent with \citet{Constante-Amores_prf_2020}).
Additionally, the escape from pinch-off causes the length of the ligament to grow over time in the $(x)$-direction, forming a secondary bulbous and a secondary neck as well as a shift in the axial curvature signs (see figure \ref{s_plot}k).

To provide conclusive evidence that Marangoni stresses drive the reopening of the neck, we run an additional simulation 
in which Marangoni stresses, or surface tension gradients, are deactivated, i.e. $\nabla_s \sigma=0$, to isolate the effects of a reduction in surface tension from those rising from Marangoni stresses. The Marangoni-suppressed case resembles the surfactant-free dynamics culminating in end-pinching (see Supplemental animation). Additionally, we performed an additional surfactant-free simulation at the ``effective" Ohnesorge number, obtained by using the surface tension achieved by the surfactant, i.e. $Oh_{eff}=\mu/\sqrt{\rho h_0 \sigma_0}=0.103$ (for $\beta_s=0.5$ and $\Gamma_0=0.5$) finding similar results. Consequently we claim Marangoni stresses are responsible of the change of interfacial dynamics discussed above.

\begin{figure}
\begin{center} 
\begin{tabular}{ccc}
\includegraphics[width=0.33\linewidth]{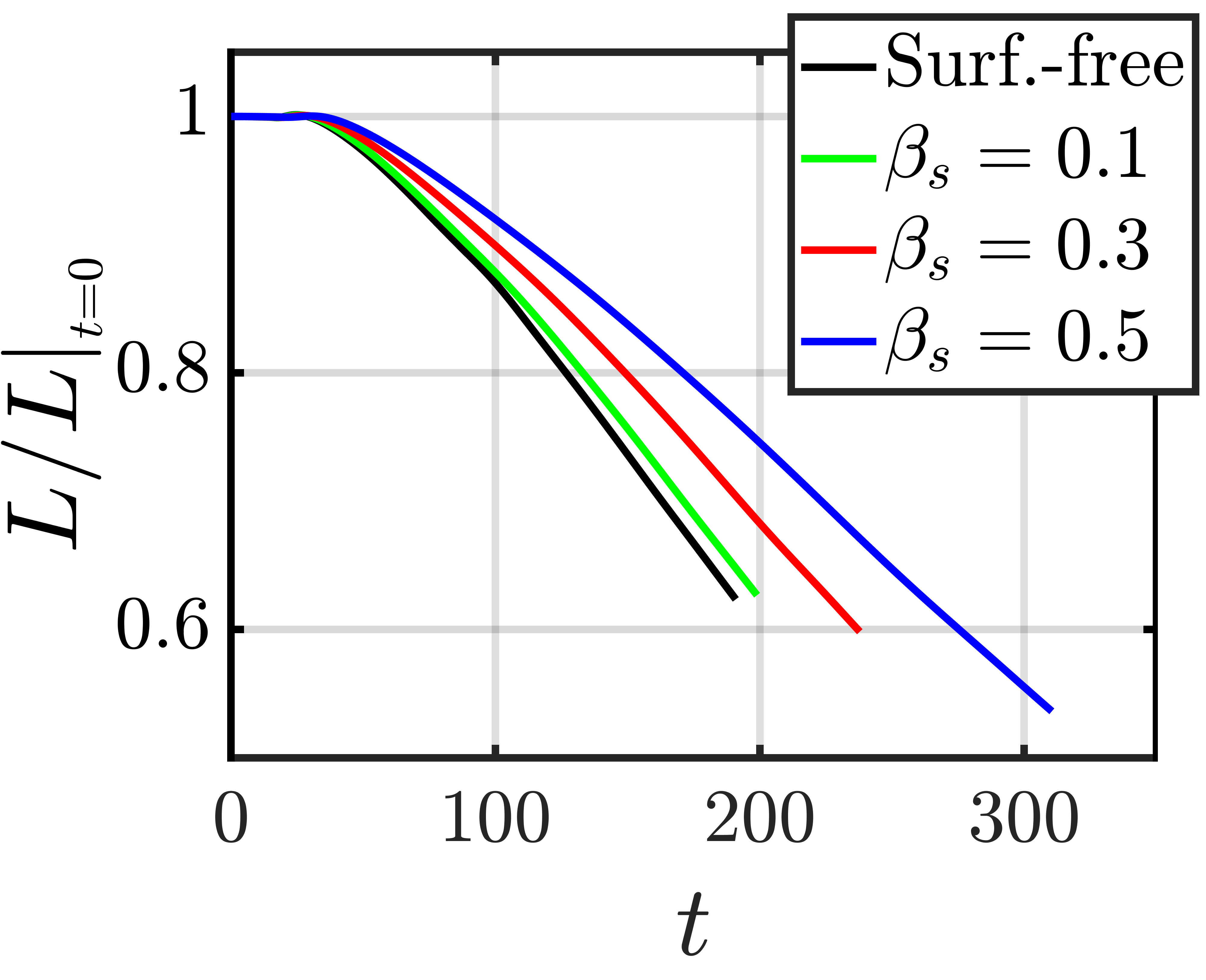}     &
\includegraphics[width=0.33\linewidth]{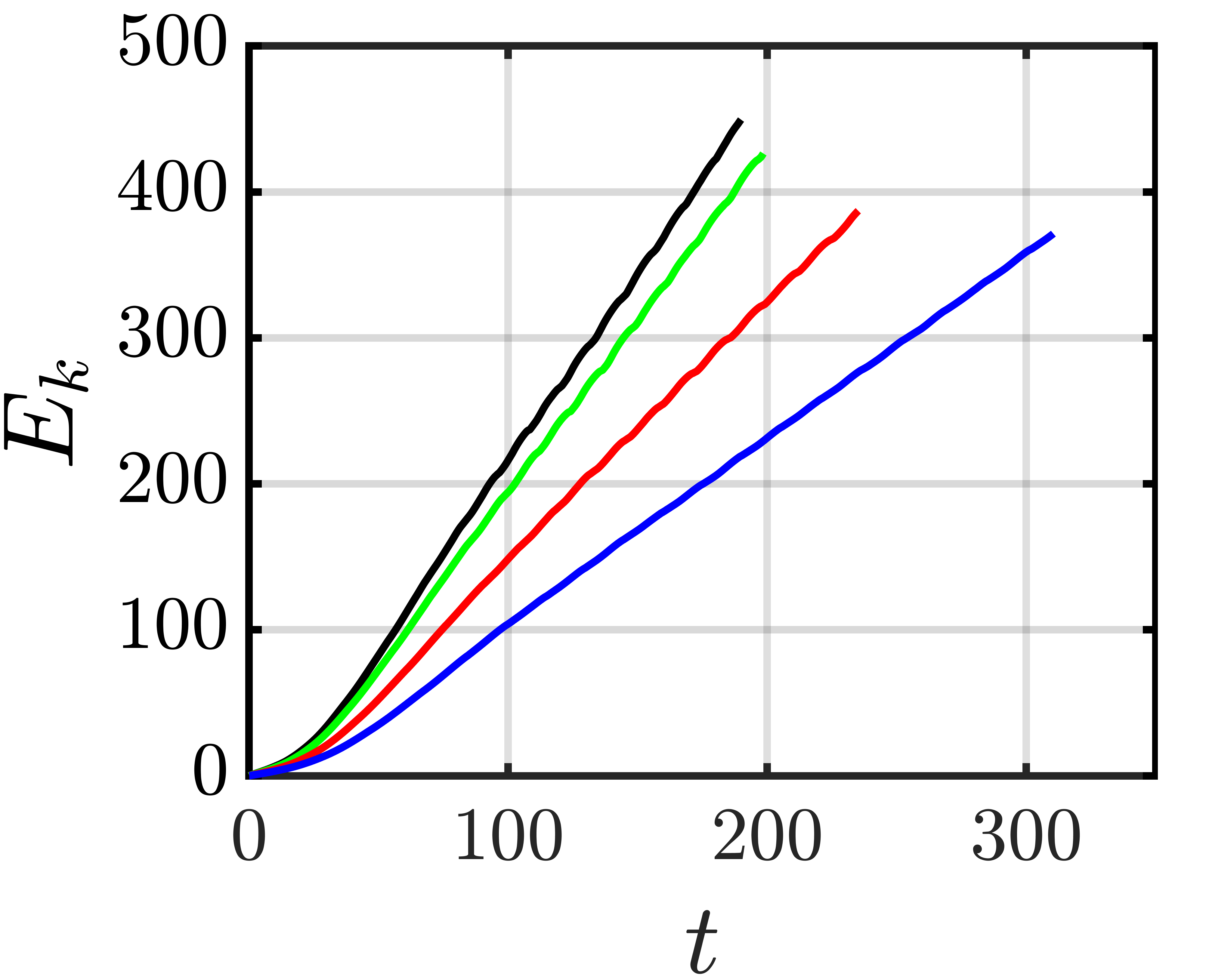}   &
\includegraphics[width=0.33\linewidth]{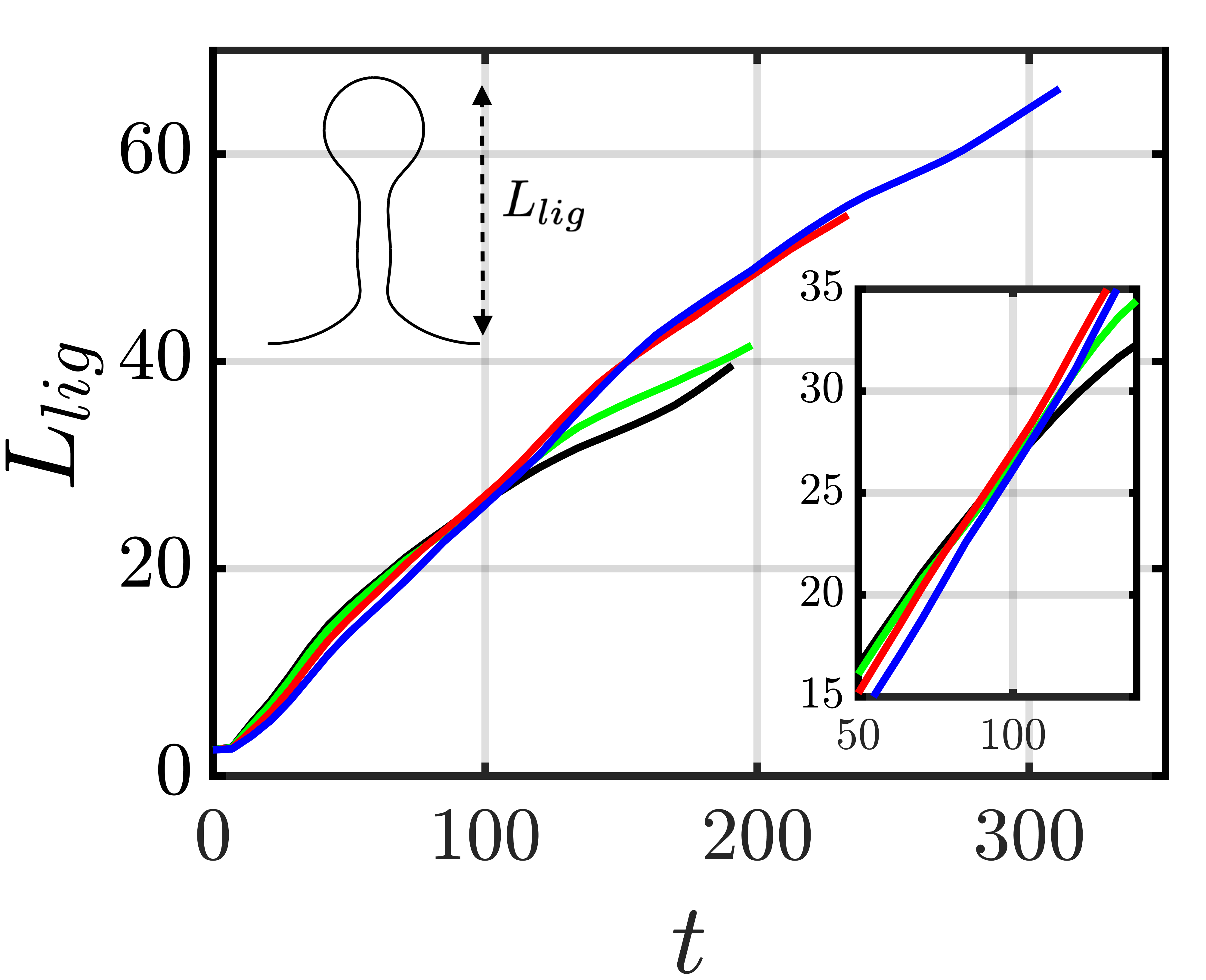} 
    \\
  (a)   & (b)  & (c)
\end{tabular}
\end{center} 
\caption{
 Temporal evolution of the ligament's tip position, the kinetic energy, and the length of the ligament, (a-c), respectively, for  the surfactant-free and surfactant-laden cases. Simulations are stopped when the sheet sheds its first drop. All parameters remain unchanged from figure \ref{temporal_evolution}. \label{metrics}} 
\end{figure}

Finally, figure \ref{metrics} shows the effect of surfactant dynamics on the  filament's tip location, $L$, the kinetic energy, $E_k= \int_{\mathcal{V}} (\rho \textbf{u}^2)/2 d{\mathcal{V}}$, and the length of the ligament (defined from the tip of the bulbous edge to its rim). Here, the kinetic energy has been normalised by the surface energy $E_s = A_0\sigma_s$, where $A_0$ is the initial surface of the sheet.
The evidence for a surfactant-driven retardation of the dynamics can be seen by inspection of $E_k$ and the streamwise location of the tip, which reveals that increasing $\beta_s$ monotonically decreases the overall value of both $E_k$ and $L$. Thus, the interfacial flow dynamics are retarded by Marangoni stresses resulting in a reduction in the retraction velocity. 
In figure \ref{metrics}c, we report the length of the ligament until pinch-off,  the addition of surfactant increases the ligament length  prior to its breakup becoming particularly pronounced at high $\beta_s$.

\section{Concluding remarks}\label{sec:Con}

Results from three-dimensional numerical simulations of the retracting dynamics of thin liquid sheets 
in the presence of insoluble surfactants were presented.
The liquid properties in the simulations were chosen to be consistent with a realistic air/water system, and the retracting velocity of the rim formed at the end is in good agreement with the Taylor-Culick speed. 
For the surfactant-laden cases, we have demonstrated that surfactant-induced Marangoni stresses drive a flow from the high surfactant concentration regions to the low concentration ones, reducing the kinetic energy, affecting the location of the bulbous tip and suppressing end-pinching.
With increasing elasticity number, the Marangoni stresses play a major role in the interfacial dynamics with the progressive elimination of the capillary wave structures 
upstream of the rim, where a complete interfacial rigidification is observed for large elasticity numbers.
Additionally, Marangoni stresses drive flow towards the sheet resulting in the  reopening of the  film thickness adjacent to the rim. 
Future research avenues  are related to examine the role of solubility on the flow dynamics. \\ 


CRC-A and AAC-P acknowledge the support from the Royal Society through a University Research Fellowship (URF/R/180016), an Enhancement Grant (RGF/EA/181002) and two NSF/CBET-EPSRC grants (Grant Nos. EP/S029966/1 and EP/W016036/1). All authors are grateful by the computing time granted by the Institut du Developpement et des Ressources en Informatique Scientifique (IDRIS) of the Centre National de la Recherche Scientifique (CNRS), coordinated by GENCI (Grand  Equipement National de Calcul Intensif) Grant No. 2022A0122B06721. Simulations were performed using code BLUE \citep{Shin_jmst_2017} and the visualisations were generated using Paraview.

Declaration of Interests. The authors report no conflict of interest.

\bibliographystyle{jfm}
\bibliography{jfm-instructions.bib}

\begin{thebibliography}{50}
\expandafter\ifx\csname natexlab\endcsname\relax\def\natexlab#1{#1}\fi
\def\au#1{#1} \def\ed#1{#1} \def\yr#1{#1}\def\at#1{#1}\def\jt#1{\textit{#1}}
  \def\bt#1{#1}\def\bvol#1{\textbf{#1}} \def\vol#1{#1} \def\pg#1{#1}
  \def\publ#1{#1}\def\arxiv#1{#1}\def\org#1{#1}\def\st#1{\textit{#1}}

\bibitem[Agbaglah(2021)]{agbaglah_2021}
{\sc \au{Agbaglah, G.G.}} \yr{2021}  \at{Breakup of thin liquid sheets through
  hole–hole and hole–rim merging}.  \jt{J. Fluid Mech.}  \bvol{911},
  \pg{A23}.

\bibitem[Agbaglah {\em et~al.\/}(2013)Agbaglah, Josserand \&
  Zaleski]{Agbaglah_2013}
{\sc \au{Agbaglah, G.}, \au{Josserand, C.} \& \au{Zaleski, S.}} \yr{2013}
  \at{Longitudinal instability of a liquid rim}.  \jt{Phys. Fluids}
  \bvol{25}~(2),  \pg{022103}.

\bibitem[Ambravaneswaran {\em et~al.\/}(2000)Ambravaneswaran, Phillips \&
  Basaran]{Ambravaneswaran}
{\sc \au{Ambravaneswaran, Bala}, \au{Phillips, Scott~D.} \& \au{Basaran,
  Osman~A.}} \yr{2000}  \at{Theoretical analysis of a dripping faucet}.
  \jt{Phys. Rev. Lett.}  \bvol{85},  \pg{5332--5335}.

\bibitem[Batchvarov {\em et~al.\/}(2020)Batchvarov, Kahouadji, Magnini,
  Constante-Amores, Craster, Shin, Chergui, Juric \&
  Matar]{batchvarov2020effect}
{\sc \au{Batchvarov, A.}, \au{Kahouadji, L.}, \au{Magnini, M.},
  \au{Constante-Amores, C.~R.}, \au{Craster, R. V~.}, \au{Shin, S.},
  \au{Chergui, J.}, \au{Juric, D.} \& \au{Matar, O.~K.}} \yr{2020}  \at{Effect
  of surfactant on elongated bubbles in capillary tubes at high reynolds
  number}.  \jt{Phys. Rev. Fluids}  \bvol{5},  \pg{093605}.

\bibitem[Bremond \& Villermaux(2006)]{bremond_villermaux_2006}
{\sc \au{Bremond, N.} \& \au{Villermaux, E.}} \yr{2006}  \at{Atomization by jet
  impact}.  \jt{J. Fluid Mech.}  \bvol{549},  \pg{273–306}.

\bibitem[Brenner \& Gueyffier(1999)]{Brenner_1999}
{\sc \au{Brenner, M.~P.} \& \au{Gueyffier, D.}} \yr{1999}  \at{On the bursting
  of viscous films}.  \jt{Phys. Fluids}  \bvol{11}~(3),  \pg{737--739}.

\bibitem[Constante-Amores {\em et~al.\/}(2021{\natexlab{{\em
  a\/}}})Constante-Amores, Batchvarov, Kahouadji, Shin, Chergui, Juric \&
  Matar]{constante_coales}
{\sc \au{Constante-Amores, C.R.}, \au{Batchvarov, A.}, \au{Kahouadji, L.},
  \au{Shin, S.}, \au{Chergui, J.}, \au{Juric, D.} \& \au{Matar, O.K.}}
  \yr{2021{\natexlab{{\em a\/}}}}  \at{Role of surfactant-induced marangoni
  stresses in drop-interface coalescence}.  \jt{J. Fluid Mech.}  \bvol{925},
  \pg{A15}.

\bibitem[Constante-Amores {\em et~al.\/}(2021{\natexlab{{\em
  b\/}}})Constante-Amores, Kahouadji, Batchvarov, Shin, Chergui, Juric \&
  Matar]{constante_jets}
{\sc \au{Constante-Amores, C.R.}, \au{Kahouadji, L.}, \au{Batchvarov, A.},
  \au{Shin, S.}, \au{Chergui, J.}, \au{Juric, D.} \& \au{Matar, O.K.}}
  \yr{2021{\natexlab{{\em b\/}}}}  \at{Direct numerical simulations of
  transient turbulent jets: vortex-interface interactions}.  \jt{J. Fluid
  Mech.}  \bvol{922},  \pg{A6}.

\bibitem[Constante-Amores {\em et~al.\/}(2020)Constante-Amores, Kahouadji,
  Batchvarov, S., Chergui, Juric \& Matar]{Constante-Amores_prf_2020}
{\sc \au{Constante-Amores, C.~R.}, \au{Kahouadji, L.}, \au{Batchvarov, A.},
  \au{S., Seungwon}, \au{Chergui, J.}, \au{Juric, D.} \& \au{Matar, O.~K.}}
  \yr{2020}  \at{Dynamics of retracting surfactant-laden ligaments at
  intermediate ohnesorge number}.  \jt{Phys. Rev. Fluids}  \bvol{5},
  \pg{084007}.

\bibitem[Constante-Amores {\em et~al.\/}(2021{\natexlab{{\em
  c\/}}})Constante-Amores, Kahouadji, Batchvarov, Shin, Chergui, Juric \&
  Matar]{constanteamores2020bb}
{\sc \au{Constante-Amores, C.~R.}, \au{Kahouadji, L.}, \au{Batchvarov, A.},
  \au{Shin, S.}, \au{Chergui, J.}, \au{Juric, D.} \& \au{Matar, O.~K.}}
  \yr{2021{\natexlab{{\em c\/}}}}  \at{Dynamics of a surfactant-laden bubble
  bursting through an interface}.  \jt{J. Fluid Mech.}  \bvol{911},  \pg{A57}.

\bibitem[Craster {\em et~al.\/}(2002)Craster, Matar \&
  Papageorgiou]{craster_pof_2002}
{\sc \au{Craster, R.~V.}, \au{Matar, O.~K.} \& \au{Papageorgiou, D.~T.}}
  \yr{2002}  \at{Pinchoff and satellite formation in surfactant covered viscous
  threads}.  \jt{Phys. Fluids}  \bvol{14}~(4),  \pg{1364--1376}.

\bibitem[Culick(1960)]{Culick}
{\sc \au{Culick, F. E.~C.}} \yr{1960}  \at{Comments on a ruptured soap film}.
  \jt{J. Appl. Phys.}  \bvol{31},  \pg{1128}.

\bibitem[Debrégeas {\em et~al.\/}(1995)Debrégeas, Martin \&
  Brochard-Wyart]{Debregeas_1995}
{\sc \au{Debrégeas, G.}, \au{Martin, P.} \& \au{Brochard-Wyart, F.}} \yr{1995}
   \at{Viscous bursting of suspended films}.  \jt{Phys. Rev. Lett.}  \bvol{75},
   \pg{3886}.

\bibitem[Driessen {\em et~al.\/}(2013)Driessen, Jeurissen, Wijshoff, Toschi \&
  Lohse]{Driessen}
{\sc \au{Driessen, T.}, \au{Jeurissen, R.}, \au{Wijshoff, H.}, \au{Toschi, F.}
  \& \au{Lohse, D.}} \yr{2013}  \at{Stability of viscous long liquid
  filaments}.  \jt{Phys. Fluids}  \bvol{25}~(6),  \pg{062109}.

\bibitem[Dupre(1867)]{Dupre_1867}
{\sc \au{Dupre, M.~A.}} \yr{1867}  \at{Sixieme memoire sur la theorie
  mechanique de la chaleur}.  \jt{Ann. Chim. Phys.}  \bvol{4}~(11),
  \pg{194–220}.

\bibitem[Fullana \& Zaleski(1999)]{Fullana_1999}
{\sc \au{Fullana, J.~M.} \& \au{Zaleski, S.}} \yr{1999}  \at{Stability of a
  growing end rim in a liquid sheet of uniform thickness}.  \jt{Phys. Fluids}
  \bvol{11},  \pg{952}.

\bibitem[Gordillo {\em et~al.\/}(2011)Gordillo, Agbaglah, Duchemin \&
  Josserand]{Gordillo_2011}
{\sc \au{Gordillo, L.}, \au{Agbaglah, G.}, \au{Duchemin, L.} \& \au{Josserand,
  C.}} \yr{2011}  \at{Asymptotic behavior of a retracting two-dimensional fluid
  sheet}.  \jt{Phys. Fluids}  \bvol{23},  \pg{122101}.

\bibitem[Harlow \& Welch(1965)]{Harlow_pof_1965}
{\sc \au{Harlow, F.~H.} \& \au{Welch, J.~E.}} \yr{1965}  \at{Numerical
  calculation of time‐dependent viscous incompressible flow of fluid with
  free surface}.  \jt{Phys. Fluids}  \bvol{8}~(12),  \pg{2182--2189}.

\bibitem[Kamat {\em et~al.\/}(2020)Kamat, Wagoner, Castrejón-Pita,
  Castrejón-Pita, Anthony \& Basaran]{kamat_2020}
{\sc \au{Kamat, P.~M.}, \au{Wagoner, B.~W.}, \au{Castrejón-Pita, A.~A.},
  \au{Castrejón-Pita, J.~R.}, \au{Anthony, C.~R.} \& \au{Basaran, O.~A.}}
  \yr{2020}  \at{Surfactant-driven escape from endpinching during contraction
  of nearly inviscid filaments}.  \jt{J. Fluid Mech.}  \bvol{899},  \pg{A28}.

\bibitem[Kamat {\em et~al.\/}(2018)Kamat, Wagoner, Thete \&
  Basaran]{Kamat_prf_2018}
{\sc \au{Kamat, P.~M.}, \au{Wagoner, B.~W.}, \au{Thete, S.~S.} \& \au{Basaran,
  O.~A.}} \yr{2018}  \at{Role of marangoni stress during breakup of
  surfactant-covered liquid threads: Reduced rates of thinning and microthread
  cascades}.  \jt{Phys. Rev. Fluids}  \bvol{3},  \pg{043602}.

\bibitem[Keller {\em et~al.\/}(1995)Keller, King \& Ting]{Keller_1995}
{\sc \au{Keller, J.~B.}, \au{King, A.} \& \au{Ting, L.}} \yr{1995}  \at{Blob
  formation}.  \jt{Phys. Fluids}  \bvol{7},  \pg{226}.

\bibitem[Krechetnikov(2010)]{Krechetnikov}
{\sc \au{Krechetnikov, R.}} \yr{2010}  \at{Stability of liquid sheet edges}.
  \jt{Phys. Fluids}  \bvol{22},  \pg{092101}.

\bibitem[Lhuissier \& Villermaux(2011)]{Lhuissier_2011}
{\sc \au{Lhuissier, H.} \& \au{Villermaux, E.}} \yr{2011}  \at{The
  destabilization of an initially thick liquid sheet edge}.  \jt{Phys. Fluids}
  \bvol{23},  \pg{091705}.

\bibitem[Liao {\em et~al.\/}(2006)Liao, Franses \& Basaran]{Liao_2006}
{\sc \au{Liao, YC}, \au{Franses, E.~I.} \& \au{Basaran, O.~A.}} \yr{2006}
  \at{Deformation and breakup of a stretching liquid bridge covered with an
  insoluble surfactant monolayer}.  \jt{Phys. Fluids}  \bvol{18}~(2),
  \pg{022101}.

\bibitem[McEntee \& Mysels(1969)]{McEntee_1969}
{\sc \au{McEntee, W.~R.} \& \au{Mysels, K.~J.}} \yr{1969}  \at{Bursting of soap
  films. i. an experimental study}.  \jt{J. Phys. Chem}  \bvol{73}~(9),
  \pg{3018--3028}.

\bibitem[Meister \& Scheele(1969)]{Meister_aiche_1969}
{\sc \au{Meister, B.~J.} \& \au{Scheele, G.~F.}} \yr{1969}  \at{Prediction of
  jet length in immiscible liquid systems}.  \jt{AIChE Journal}  \bvol{15}~(5),
   \pg{689--699}.

\bibitem[Mirjalili {\em et~al.\/}(2018)Mirjalili, Chan \&
  Mani]{Mirjalili_snh_2018}
{\sc \au{Mirjalili, S.}, \au{Chan, W. H.~R.} \& \au{Mani, A.}} \yr{2018}
  \at{High fidelity simulations of micro-bubble shedding from retracting thin
  gas films in the context of liquid-liquid impact}.  \jt{In 32nd Symposium on
  Naval Hydrodynamics} .

\bibitem[Peskin(1977)]{Peskin_jcp_1977}
{\sc \au{Peskin, C.~S}} \yr{1977}  \at{Numerical analysis of blood flow in the
  heart}.  \jt{J. ~Comput. ~ Phys}  \bvol{25}~(3),  \pg{220 -- 252}.

\bibitem[Ranz(1959)]{Ranz}
{\sc \au{Ranz, W.~E.}} \yr{1959}  \at{Some experiments on the dynamics of
  liquid films}.  \jt{Int. J. Appl. Phys.}  \bvol{30}~(12),  \pg{1950--1955}.

\bibitem[Rayleigh(1879)]{Rayleigh_1879}
{\sc \au{Rayleigh, L.}} \yr{1879}  \at{On the capillary phenomena of jets}.
  \jt{Proc. R. Soc. London}  \bvol{29},  \pg{71--97}.

\bibitem[Rayleigh(1891)]{Rayleigh_1891}
{\sc \au{Rayleigh, L.}} \yr{1891}  \at{Some applications of photography}.
  \jt{Nature}  \bvol{44}~(1133),  \pg{249--254}.

\bibitem[Rieber \& Frohn(1999)]{Rieber}
{\sc \au{Rieber, M.} \& \au{Frohn, A.}} \yr{1999}  \at{A numerical study on the
  mechanism of splashing}.  \jt{Int J Heat Fluid}  \bvol{20}~(5),
  \pg{455--461}.

\bibitem[Roisman {\em et~al.\/}(2006)Roisman, K.Horvat \& Tropea]{Roisman}
{\sc \au{Roisman, I.~V.}, \au{K.Horvat} \& \au{Tropea, C.}} \yr{2006}
  \at{Spray impact: Rim transverse instability initiating fingering and splash,
  and description of a secondary spray}.  \jt{Phys. Fluids}  \bvol{18},
  \pg{102104}.

\bibitem[Savva \& Bush(2009)]{savva_bush_2009}
{\sc \au{Savva, N.} \& \au{Bush, J. W.~M.}} \yr{2009}  \at{Viscous sheet
  retraction}.  \jt{J. Fluid Mech.}  \bvol{626},  \pg{211–240}.

\bibitem[Shin {\em et~al.\/}(2017)Shin, Chergui \& Juric]{Shin_jmst_2017}
{\sc \au{Shin, S.}, \au{Chergui, J.} \& \au{Juric, D.}} \yr{2017}  \at{A solver
  for massively parallel direct numerical simulation of three-dimensional
  multiphase flows}.  \jt{J. Mech. Sci. Tech.}  \bvol{31},  \pg{1739--1751}.

\bibitem[Shin {\em et~al.\/}(2018)Shin, Chergui, Juric, Kahouadji, Matar \&
  Craster]{Shin_jcp_2018}
{\sc \au{Shin, S.}, \au{Chergui, J.}, \au{Juric, D.}, \au{Kahouadji, L.},
  \au{Matar, O.~K.} \& \au{Craster, R.~V.}} \yr{2018}  \at{A hybrid interface
  tracking -- level set technique for multiphase flow with soluble surfactant}.
   \jt{J. Comp. Phys.}  \bvol{359},  \pg{409--435}.

\bibitem[Shin \& Juric(2002)]{Shin_jcp_2002}
{\sc \au{Shin, S.} \& \au{Juric, D.}} \yr{2002}  \at{Modeling three-dimensional
  multiphase flow using a level contour reconstruction method for front
  tracking without connectivity}.  \jt{J. Comput. Phys}  \bvol{180},
  \pg{427--470}.

\bibitem[Shin \& Juric(2009)]{Shin_ijnmf_2009}
{\sc \au{Shin, S.} \& \au{Juric, D.}} \yr{2009}  \at{A hybrid interface method
  for three-dimensional multiphase flows based on front-tracking and level set
  techniques}.  \jt{Int. J. Num. Meth. Fluids}  \bvol{60},  \pg{753--778}.

\bibitem[Song \& Tryggvason(1999)]{Song_1999}
{\sc \au{Song, M.} \& \au{Tryggvason, G.}} \yr{1999}  \at{The formation of
  thick borders on an initially stationary fluid sheet}.  \jt{Phys. Fluids}
  \bvol{11}~(9),  \pg{2487--2493}.

\bibitem[Sussman {\em et~al.\/}(1994)Sussman, Smereka \&
  Osher]{Sussman_cp_1998}
{\sc \au{Sussman, M.}, \au{Smereka, P.} \& \au{Osher, S.}} \yr{1994}  \at{A
  level set approach for computing solutions to incompressible two-phase flow}.
   \jt{J. Comp. Phys.}  \bvol{114}~(1),  \pg{146 -- 159}.

\bibitem[Sünderhauf {\em et~al.\/}(2002)Sünderhauf, Raszillier \&
  F.]{Sunderhauf_2002}
{\sc \au{Sünderhauf, G.}, \au{Raszillier, H.} \& \au{F., Durst}} \yr{2002}
  \at{The retraction of the edge of a planar liquid sheet}.  \jt{Phys. Fluids}
  \bvol{14},  \pg{198}.

\bibitem[Taylor(1959)]{Taylor_1959}
{\sc \au{Taylor, G.~I.}} \yr{1959}  \at{The dynamics of thin sheets of fluid.
  ii. waves on fluid sheets}.  \jt{Proc. R. Soc. London A}  \bvol{253},
  \pg{296}.

\bibitem[Timmermans \& Lister(2002)]{timmermans_lister_2002}
{\sc \au{Timmermans, M-L~E.} \& \au{Lister, J.~R.}} \yr{2002}  \at{The effect
  of surfactant on the stability of a liquid thread}.  \jt{J. Fluid Mech.}
  \bvol{459},  \pg{289–306}.

\bibitem[Villermaux(2020)]{villermaux_2020}
{\sc \au{Villermaux, E.}} \yr{2020}  \at{Fragmentation versus cohesion}.
  \jt{J. Fluid Mech.}  \bvol{898},  \pg{P1}.

\bibitem[Villermaux \& Bossa(2011)]{Villermaux_2011}
{\sc \au{Villermaux, E.} \& \au{Bossa, B.}} \yr{2011}  \at{Drop fragmentation
  on impact}.  \jt{J. Fluid Mech.}  \bvol{668},  \pg{412}.

\bibitem[Wang \& Bourouiba(2018)]{wang_bourouiba_2018}
{\sc \au{Wang, Y.} \& \au{Bourouiba, L.}} \yr{2018}  \at{Unsteady sheet
  fragmentation: droplet sizes and speeds}.  \jt{J. Fluid Mech.}  \bvol{848},
  \pg{946–967}.

\bibitem[Wang \& Bourouiba(2021)]{wang_bourouiba_2021}
{\sc \au{Wang, Y.} \& \au{Bourouiba, L.}} \yr{2021}  \at{Growth and breakup of
  ligaments in unsteady fragmentation}.  \jt{J. Fluid Mech.}  \bvol{910},
  \pg{A39}.

\bibitem[Wang {\em et~al.\/}(2018)Wang, Dandekar, Bustos, Poulain \&
  Bourouiba]{Wang_prl}
{\sc \au{Wang, Y.}, \au{Dandekar, R.}, \au{Bustos, N.}, \au{Poulain, S.} \&
  \au{Bourouiba, L.}} \yr{2018}  \at{Universal rim thickness in unsteady sheet
  fragmentation}.  \jt{Phys. Rev. Lett.}  \bvol{120},  \pg{204503}.

\bibitem[Yarin \& Weiss(1995)]{yarin_weiss_1995}
{\sc \au{Yarin, A.~L.} \& \au{Weiss, D.~A.}} \yr{1995}  \at{Impact of drops on
  solid surfaces: self-similar capillary waves, and splashing as a new type of
  kinematic discontinuity}.  \jt{J. Fluid Mech.}  \bvol{283},  \pg{141–173}.

\bibitem[Zhang {\em et~al.\/}(2010)Zhang, Brunet, Eggers \& Deegan]{Zhang_2010}
{\sc \au{Zhang, L.~V.}, \au{Brunet, P.}, \au{Eggers, J.} \& \au{Deegan, R.~D.}}
  \yr{2010}  \at{Wavelength selection in the crown splash}.  \jt{Phys. Fluids}
  \bvol{22},  \pg{122105}.

\end{thebibliography}

\end{document}